\documentclass[aps,prc,twocolumn,showpacs,preprintnumbers,floatfix,nofootinbib,amsmath,amssymb,balancelastpage,table]{revtex4-1}
\usepackage{graphicx, color, dcolumn, bm, amsmath}
\usepackage{multirow}
\usepackage{url}

\renewcommand{\epsilon}{\varepsilon}

\newcommand{\str}{{\eta_\mathrm{s}}}

\newcommand{\BV[1]}{\mbox{\boldmath ${#1}$}}

\begin{document}
\preprint{RIKEN-QHP-204}
\title{Interplay between Mach cone and radial expansion and its signal in $\gamma$-jet events}

\author{Yasuki~Tachibana}
\email{yasuki.tachibana@mail.ccnu.edu.cn}

\affiliation{
Theoretical Research Division, Nishina Center, RIKEN, Wako 351-0198, Japan
}
\affiliation{
Department of Physics, Sophia University, Tokyo 102-8554, Japan
}

\author{Tetsufumi~Hirano}
\affiliation{
Department of Physics, Sophia University, Tokyo 102-8554, Japan
}

\date{\today}

\begin{abstract}
We study the hydrodynamic response 
to jet quenching 
in expanding quark-gluon plasma (QGP) 
and  
its signal 
in the resulting particle distribution. 
The ideal hydrodynamic simulations of 
the $\gamma$-jet events in heavy-ion collisions 
are performed 
in a full (3 + 1)-dimensional setup. 
The jet-induced Mach cone 
and 
the radial expansion of the background 
mutually push and distort each other. 
As a result, 
the particle emission 
is suppressed 
in the direction 
in which the radial flow is pushed back by the Mach cone 
when the jet path is an off-central one. 
This is the direct signal of hydrodynamic response to the jet
and, moreover, includes information about the jet path 
in the expanding QGP fluid. 
 \end{abstract}
\maketitle

\section{Introduction}

In heavy-ion collisions 
at the BNL Relativistic Heavy Ion Collider (RHIC) 
and 
at the CERN Large Hadron Collider (LHC), 
a bulk matter 
consisting of deconfined quarks and gluons, 
so-called quark-gluon plasma (QGP), 
is created. 
One of the key properties of QGP 
is 
fluid-dynamical behavior 
which 
is well 
described by relativistic hydrodynamics 
\cite{Heinz:2001xi, sQGP1, sQGP2, sQGP3,Hirano:2005wx, 
Schenke:2010rr,Qiu:2011iv, Qiu:2011hf, Gale:2012rq}. 
The fluidity implies strong coupling among 
the constituents of QGP. 
At the same time, 
QGP has a large stopping power against the propagation of 
energetic partons produced in the initial hard scatterings. 
These partons do not take part in the hydrodynamic bulk evolution 
and are subject to traverse the QGP medium 
because of their large transverse momenta. 
During the propagation, 
jets deposit their energies and momenta 
via the strong interaction with the QGP medium. 
This phenomenon is called jet quenching 
\cite{Bjorken:1982tu, Appel:1985dq, Blaizot:1986ma, Rammerstorfer:1990js, Gyulassy:1990ye, Thoma:1990fm,Gyulassy:1993hr}. 
As a consequence of the jet quenching, 
jet yields in heavy-ion collisions are suppressed 
compared to that in proton-proton collisions 
where the QGP medium is not produced 
\cite{Aad:2014bxa, Adam:2015ewa, CMS:2012kxa}. 
In recent heavy-ion collision experiments at LHC, 
the large center-of-mass energy made available 
detailed measurements of fully reconstructed jets 
with large transverse momenta $p_{T, {\rm jet}} >$ 100 GeV/$c$ 
\cite{Aad:2010bu, Chatrchyan:2011sx, Chatrchyan:2012nia, Chatrchyan:2012gw, CMS:2012kxa, Chatrchyan:2012gt, Aad:2012vca,  Aad:2013sla, Chatrchyan:2013kwa, Chatrchyan:2014ava, Aad:2014wha, Aad:2014bxa,  Adam:2015ewa, Aad:2015bsa, Khachatryan:2015lha, Khachatryan:2016erx}. 
The full jet evolution in the QGP has been actively studied also theoretically 
\cite{Vitev:2009rd,Qin:2010mn,CasalderreySolana:2010eh,Young:2011qx, He:2011pd, Renk:2012cx, Ma:2013pha,Senzel:2013dta,Chien:2015hda, Dai:2012am,Wang:2013cia, Qin:2012gp,MehtarTani:2011tz,CasalderreySolana:2012ef,Blaizot:2013hx,Fister:2014zxa,Apolinario:2012cg,Zapp:2012ak,Majumder:2013re, Casalderrey-Solana:2014bpa, Blaizot:2014ula}.

For QGP, 
the energetic light-quarks and gluons 
can be considered 
supersonic moving sources of 
the energy and momentum. 
Assuming that 
QGP responds hydrodynamically 
to the jet, 
the Mach cone, 
namely 
the conical shock wave, 
is excited 
when the deposited energy and momentum 
diffuse inside the medium. 
In early studies, 
motivated by 
the double-hump structure in the away-side of azimuthal correlations 
\cite{Adams:2005ph, Adler:2005ee, Adare:2008ae,Abelev:2008ac} 
as the signal of the Mach cone, 
many theoretical studies concerning the medium response to the jet were conducted to explain it 
\cite{Stoecker:2004qu, CasalderreySolana:2004qm, Ruppert:2005uz, Satarov:2005mv, Renk:2005si, Ma:2006fm, 
Chaudhuri:2006qk, Zhang:2007qx, Chaudhuri:2007vc, Li:2009ax, Betz:2010qh, Ma:2010dv}. 
However, it has turned out that the double-hump structure is well reproduced 
by the contribution of the anisotropic flow 
originating from the initial geometrical fluctuation of the nucleons in the heavy ions 
rather than by that of the Mach cone
\cite{Takahashi:2009na, Ma:2010dv, Alver:2010gr, ALICE:2011ab,Schenke:2010rr}. 
Recently, the hydrodynamic response to the jet quenching 
was focused on again 
because 
it affects 
various observables 
in detailed measurements with full jet reconstruction. 
The enhancement of low-$p_T$ particles away from the quenched jets 
\cite{Chatrchyan:2011sx, Khachatryan:2015lha} 
can be interpreted 
as a consequence of 
the energy-momentum transport by 
the Mach cone \cite{Tachibana:2014lja}. 
The soft particles from the 
medium excitation 
affect 
also 
the fragmentation functions 
and 
the jet transverse profile 
\cite{Wang:2013cia, He:2015pra}.
A flow flux 
behind the propagating parton 
is generated by the 
momentum deposition 
and 
fills in the Mach cone. 
The flow  
following the energetic partons 
is called the diffusion wake 
and 
can significantly modify 
the jet structure inside the jet cone. 

As well as the formation of a Mach cone is 
a clear manifestation of the fluidity of QGP, 
its structure is 
characterized by key properties of QGP, 
e.g., sound velocity, viscosity, and stopping power. 
The study of this 
unique phenomenon involving a shock wave 
could provide us a great opportunity 
to extract the property of QGP. 
However, the observables in heavy-ion collisions 
are the particle spectra 
at the final state of the event 
and 
the developing Mach cone itself 
cannot be seen directly. 
Therefore, 
it is necessary 
to understand 
how the signal of 
the hydrodynamic response to the jet propagation 
appears 
in the consequent particle distribution. 

In this paper, 
we study 
the hydrodynamic response 
to jet quenching 
in the QGP fluid. 
The background QGP fluid in heavy-ion collisions 
expands at relativistic flow velocity. 
The shape of the Mach cone 
is distorted by the background expansion, 
which affects significantly 
the resulting particle distribution 
\cite{Satarov:2005mv, Chaudhuri:2006qk, Chaudhuri:2007vc, Li:2010ts, Betz:2010qh, Bouras:2014rea, Tachibana:2014lja}. 
It should be also noted 
that the Mach cone has three-dimensional structure 
and violates 
the boost invariance 
that the medium profile 
is supposed to have as an approximate symmetry 
around midrapidity.
We describe 
the whole dynamics of the medium 
including 
the interplay between 
the response to jet quenching and the background expansion 
by numerically solving 
the $(3 + 1)$-dimensional 
relativistic hydrodynamic equations 
with source terms. 
We perform 
simulations of 
$\gamma$-jet events in central Pb-Pb collisions at the LHC energy 
and 
see 
how the Mach cone develops in the {\it expanding} medium. 
Then 
we investigate 
how the effect of the hydrodynamic response to jet quenching 
can be seen in the particle distribution 
after the hydrodynamic evolution. 
We show that 
the particle production 
is suppressed 
in a certain direction 
depending on the jet path in the medium 
as a result of 
the interplay between 
the Mach cone and the radial expansion. 
We also show that 
this feature 
can be clearly seen 
even in the event-averaged azimuthal angle distribution 
by introducing the trigger bias 
for the transverse momenta of the jet and the photon. 

The paper is organized as follows.
First, we present 
the formulation of the model employed in this work in Sec. \ref{sec:an}. 
In Sec. \ref{sec:sim}, 
we present the results of the simulations of 
$\gamma$-jet events in the heavy-ion collisions. 
Section \ref{sec:sum} is devoted to summary and concluding remarks. 
In the following, 
the collision axis is taken as the $z$-axis 
and 
the incoming Pb nuclei collide 
at $(t,\,z)=(0,\,0)$. 
We also use 
the Milne coordinates 
$\left(\tau, x, y, \eta_{\rm s}\right)$, 
where 
$\tau=\left(t^2-z^2\right)^{1/2}$ is the proper time 
and 
$\eta_{\rm s}=\left(1/2\right)\ln\left[\left(t+z\right)/\left(t-z\right)\right]$ is 
the spacetime rapidity. 

\section{METHOD OF ANALYSIS}\label{sec:an}

\subsection{Hydrodynamic equation with source terms}
Through a strong interaction 
with the medium, 
energetic partons lose their energy and momentum. 
We assume that 
the lost energy and momentum 
are 
instantaneously thermalized 
due to the strong interaction
and 
are transmitted to QGP evolving as a fluid. 
The equation of motion of the fluid 
with incoming energy and momentum 
is given by 
\begin{eqnarray}
 \partial_\mu T^{\mu\nu}\left(x\right)=J^{\nu}\left(x\right), \label{eqn:hws}
\end{eqnarray}
where $T^{\mu\nu}$ is the energy-momentum tensor of the fluid 
and 
$J^{\nu}$ is the source term which is 
the four-momentum density 
incoming from traversing partons to the fluid. 
Here, 
the medium is 
modeled as 
an ideal fluid, 
whose energy-momentum tensor 
can be decomposed as 
\begin{eqnarray}
 T^{\mu\nu}=\left(\epsilon+p\right)u^{\mu}u^{\nu}-p\eta^{\mu\nu}, 
\end{eqnarray}
where $\epsilon$ is the energy density, 
$p$ is the pressure, 
$u^{\mu}$ is the flow four-velocity, 
and $\eta^{\mu\nu}={\rm diag}\left(1,-1,-1,-1\right)$ 
is the Minkowski metric. 
Here, we employ a simple form 
of the source term 
for a jet traversing the QGP fluid: 
\begin{eqnarray}
J^{\mu}\left(x\right)&=&-
\frac{dp_{\rm jet}^\mu}{dt} 
\delta^{(3)}\left(\mbox{\boldmath $x$}-\mbox{\boldmath $x$}_{\rm jet}(t)\right).\label{eqn:source_tm}
\end{eqnarray}
We transform the hydrodynamic equations (\ref{eqn:hws}) to the ones 
in the Milne coordinate by performing the Lorentz transformation. 
The space-time evolution 
of the medium created 
in heavy-ion collisions 
is described 
by solving the equations in the Milne coordinate numerically. 
As an equation of state 
needed to close the system of equations, 
we employ 
the one from recent lattice {QCD} calculations 
\cite{Borsanyi:2013cga}. 
Because Eq.~(\ref{eqn:hws}) can describe 
both the hydrodynamic response to jet propagation 
and the expansion of the QGP, 
the interplay between them is automatically included 
in this framework. 

\subsection{Energy loss}
We model partons traveling through QGP as massless particles 
and neglect their structure. 
The jet particles move at the speed of light 
and deposit their energy and momentum 
via the strong interaction with the medium. 
We use the jet energy-loss of the form \cite{Betz:2010qh}, 
\begin{eqnarray}
\frac{dp_a^0}{dt} = - \left[\frac{T\left(t, \BV[x]_a\left(t\right)\right)}
{T_0}\right]^3\left.\frac{dE}{dl}\right|_0, \label{Eq:energyloss}
\end{eqnarray}
where $T_0$ is the reference temperature 
and $\left.dE/dl\right|_0$ is 
the energy-loss rate at $T_0$. 
In this study, 
the jet trajectories are restricted 
to be straight in the transverse plane. 
The jets lose their energy
when 
the jet particles 
penetrate 
the QGP medium with the local temperature above 
$160\,{\rm MeV}$. 
Here we set $T_0=500\,{\rm MeV}$
and $\left.{dE}/{dl}\right|_0=15\,{\rm GeV/fm}$, 
which reproduces the typical values of 
the nuclear modification factor 
for jets 
around $p_{T,\,\rm jet}\sim 100\,{\rm GeV}/c$ 
in heavy-ion collisions 
at LHC \cite{Aad:2014bxa}. 

\subsection{Initial profile of the medium}
A hydrodynamic description 
is applied 
to the space-time evolution of the medium 
after the thermalization 
$\tau\geq\tau_0$. 
We set the initial condition for 
the entropy density profile of the medium 
at $\tau_0=0.6$ fm/$c$ as 
\begin{eqnarray}
s\!\left(\tau_0,\!\BV[r]_{\perp},\str\!\right) 
&=&
s_T\!\left(\BV[r]_{\perp}\right)
H\!\left(\eta_{\rm s} \right)
\theta \!\left(Y_{\rm beam}\!-\!\left|\str\right|\right)
.\label{eq:modelinit0}
\end{eqnarray}
Here, 
$Y_{\rm beam}$ 
is the beam rapidity of incoming nuclei, and 
$H$ represents the profile in the $\str$ direction. 
$s_T$ is the transverse profile of 
the initial entropy density 
at midrapidity, 
\begin{eqnarray}
s_T\left(\BV[r]_{\perp}\right) 
=
\frac{C}{\tau_0}\,
\left[
\frac{\left(1-\alpha\right)}{2} 
n_{\rm part}\left(\BV[r]_\perp\right)
+\alpha n_{\rm coll}\left(\BV[r]_\perp\right)
\right],\label{eq:sT}
\end{eqnarray}
where 
$n_{\rm coll}$ 
and 
$n_{\rm part}$ 
are 
the number density 
of nucleon-nucleon binary collisions 
and 
of participating nucleons 
calculated from 
the optical Glauber model, 
respectively. 
For Pb-Pb collision at LHC, 
the parameters 
$C=19.8$ and $\alpha=0.14$ 
are 
fitted to reproduce 
the centrality dependence of multiplicity 
measured by the ALICE Collaboration 
\cite{Hirano:2012yy, Aamodt:2010cz}.

The initial profile around the midrapidity region 
is flat in the $\eta_{\rm s}$ direction, 
like the Bjorken scaling solution \cite{Bjorken:1982qr}. 
The flat region is smoothly connected to a vacuum at both ends 
by using half Gaussians: 
\begin{eqnarray}
H\!\left(\!\eta_{\rm s}\! \right)\!&=&\!
\exp\!\left[\!-\frac{\left(\left|\eta_{\rm s}\right|\!-\!\eta_{\rm flat}/2\right)^2} {2\sigma_{\eta}^2} \theta\! \left(\!\left|\eta_{\rm s}\right|\!-\!\frac{\eta_{\rm flat}}{2} \!\right) \!\right]\!,  
\label{eq:modelinit}
\end{eqnarray}
where the parameters $\eta_{\rm flat}=3.8$ and $\sigma_\eta=3.2$
are chosen to give 
the pseudorapidity distribution 
similar to 
the one obtained from 
the Monte Carlo Kharzeev-Levin-Nardi model 
for central Pb-Pb collisions \cite{Hirano:2010je}. 
It is assumed that 
there is no transverse flow 
at $\tau=\tau_0$ 
and 
the radial expansion of the medium 
is driven solely by 
the initial pressure gradient in the transverse direction. 
For the initial condition of the flow velocity in the longitudinal direction, 
the space-time rapidity component is set to zero: $u^{\eta_{\rm s}}(\tau=\tau_0)=0$ \cite{Bjorken:1982qr}. 
These initial conditions for the flow 
are commonly used in hydrodynamic models for heavy-ion collisions. 
In this study, we neglect 
the effect of the initial geometrical fluctuation of the nucleons in colliding nuclei 
and employ the smooth averaged profile. 
The flow driven by the initial fluctuation 
can modify the development of the Mach cone 
in the medium, which will be considered in upcoming work. 

\subsection{Freeze-out}
To obtain the particle spectrum, 
we switch 
from 
the hydrodynamic description 
to a phase-space distribution of individual particles 
via 
the Cooper-Frye formula \cite{Cooper:1974mv}, 
\begin{eqnarray}
\frac{dN}{p_Tdp_Td\phi_pd\eta}
\!=\!\sum_i\!
\frac{d_i}{\left(2\pi\right)^3}\!\!\int_{\!\Sigma}\!
\frac
{p^\mu d\sigma_{\mu}\left(x\right)}
{\exp\!\left[{p^{\mu}u_{\mu}\left(x\right)}/{T\left(x\right)}\right]\!\mp_{\rm BF}\!1}. \label{eqn:C-F}
\end{eqnarray}
Here, 
$p_T$ is the transverse momentum, 
$\phi_p$ is the azimuthal angle, 
$\eta$ is the rapidity, and 
$\Sigma$ is 
the freeze-out hypersurface 
which is 
determined 
by assuming 
the isothermal freeze-out 
at the freeze-out temperature $T_f$. 
$d_i$ is the degeneracy and $\mp_{\rm BF}$ corresponds to 
Bose or Fermi statistics for particle species $i$. 
We set $T_f=145\,{\rm MeV}$, 
which is the typical value 
to obtain 
the observed $p_T$ spectra 
and not crucial for results presented here. 
In this study, 
we calculate 
the distribution of charged pions 
directly emitted from 
the freeze-out hypersurface 
and 
investigate how the effect of jet-induced flow 
appears in it. 
The charged pions from decays of hadron resonances 
after the hydrodynamic evolution are not included here. 
For more quantitative analysis, their contribution should be considered 
and are left as a subject for future study. 
Since the azimuthal angle distributions of the hadron resonances 
are expected to be similar to that of the charged pions 
just after the hydrodynamic evolution, 
the contribution from the decays 
can increase the amplitude of the resulting distribution. 

\section{Simulations and Results}\label{sec:sim}
We perform simulations of the events 
in which one jet particle travels through the QGP medium. 
These correspond to $\gamma$-jet events 
originating from the pair production of 
one photon and one parton at initial hard scatterings. 
At leading order, 
the photon and the parton 
having the same initial energies 
propagate in opposite directions. 
While the parton 
deposits its energy and momentum, 
the photon freely penetrates 
the medium. 

Actually, 
the energies of the observed jet and photon 
are not exactly the same 
even in proton-proton collisions due to higher order contributions. 
The momentum fraction of a photon tagged jet, 
$x_T=p_{T,{\rm jet}}/p_{T,{\rm \gamma}}$, 
is distributed with a finite width around a peak at $x_T=1$. 
In $\gamma$-jet events in heavy-ion collisions, 
the energy loss of jets is observed 
as a shift of the peak to less than unity 
\cite{Chatrchyan:2012gt}. 
In this study, 
we neglect 
such higher order contributions 
and set 
the initial energies of 
the parton and photon to be the same in each event. 
The parton is regarded as a jet and 
travels through the medium. 
Furthermore, we also do not consider the structure of the jet. 
In reality, however, jets in heavy-ion collisions 
radiate gluons due to both 
their high virtuality (vacuum cascade) 
and inelastic collisions with the medium constituents (in-medium cascade). 
The jets have shower structures evolving during their propagation. 
It should be noted that , in particular, the 
spatial profile of jet energy deposition can affect 
the pattern of the medium response. 
In the case where 
the radiated gluons are emitted at small angles, 
the jet 
has a collimated structure 
and 
induces a Mach cone similar to the one in the case of one parton 
\cite{Neufeld:2009ep, Qin:2009uh, Neufeld:2011yh}. 
However, when 
the radiated gluons are emitted at large angles 
with sufficiently large momenta, 
the jet has a widespread structure 
and 
each radiated gluon becomes 
a separate source of a Mach cone. 
As a result, 
the medium response to the jet 
becomes a superposition of the distinct Mach cones 
and the clear conical structure 
cannot be seen \cite{Neufeld:2011yh}. 
In this study, 
we assume that the gluons having sufficiently large momenta are radiated only at small angle 
and regard them as a part of the structureless jet. To perform more realistic simulations, 
the inclusion of 
the proper $x_T$ distribution 
and 
the jet structure evolution 
is necessary 
(as in Ref. \cite{Qin:2012gp}). 
Here we postpone it as a future study. 

The shape of the medium response to the jet 
is also affected by the viscosity of the QGP fluid. 
As the shear viscosity increases, it tends to smear out the Mach cone structure 
because the diffusion wake is transported perpendicular to the jet 
and destroys the conical wave front 
\cite{Neufeld:2008dx, Neufeld:2010tz, Bouras:2012mh, Bouras:2014rea}. 
Nevertheless, 
the rather clear Mach cone structure appears in 
the calculation with the small shear viscosity 
which the QGP is expected to have. 
Here we neglect the small shear viscosity of the QGP 
and model the QGP as the ideal fluid. 
In this sense, the Mach cone develops with the most clear conical structure 
and its maximum signal is exhibited in our results. 

In the simulations, 
we consider 
the completely central Pb-Pb collisions 
(impact parameter $b = 0\,{\rm fm}$). 
The profile of the created medium at $\tau_0$ 
is 
isotropic in the transverse plane and  
the origin 
$\left(x,\,y,\,\eta_{\rm s}\right)=\left(0\,{\rm fm},\,0\,{\rm fm},\,0\right)$ 
is set at its center. 
The jet parton is supposed 
to be created 
in the transverse plane $\eta_{\rm s}=0$ 
at $\tau=0$ 
and 
travel freely without any interaction 
until $\tau_0=0.6$ fm/$c$. 
Then 
it starts to 
interact with 
the QGP 
at the same time as the beginning of 
the hydrodynamic bulk evolution. 
The space-time evolution of 
the medium with 
the incoming energy via 
the jet energy loss 
is assumed to obey Eq. (\ref{eqn:hws}). 
Without a loss of generality, 
we set the direction of the photon propagation 
as the positive $x$ direction 
$\left(\phi_p=0,\eta=0\right)$ 
and the direction of the jet propagation 
as the negative $x$ direction 
$\left(\phi_p=\pi,\eta=0\right)$. 
In the completely central collisions, 
any path of the jets in the QGP is covered 
by changing only 
the jet production point due to rotational symmetry in the transverse plane. 

\subsection{Energy density distribution of the medium}
\begin{figure*}
  \begin{center}
   \includegraphics[width=5.4cm,bb=0 0 500 500]{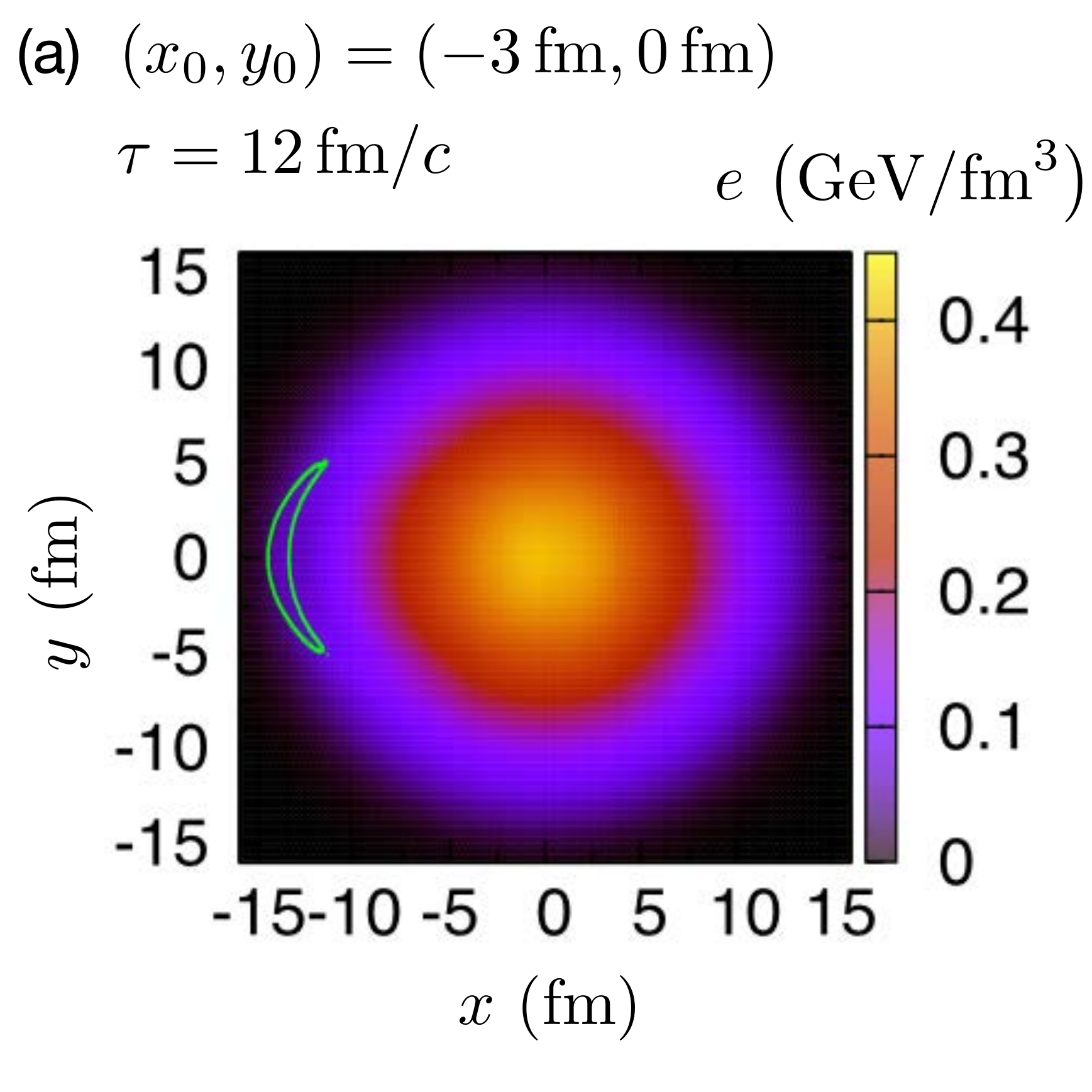}
  \hspace{14pt}
  \vspace{4pt}
   \includegraphics[width=5.4cm,bb=0 0 500 500]{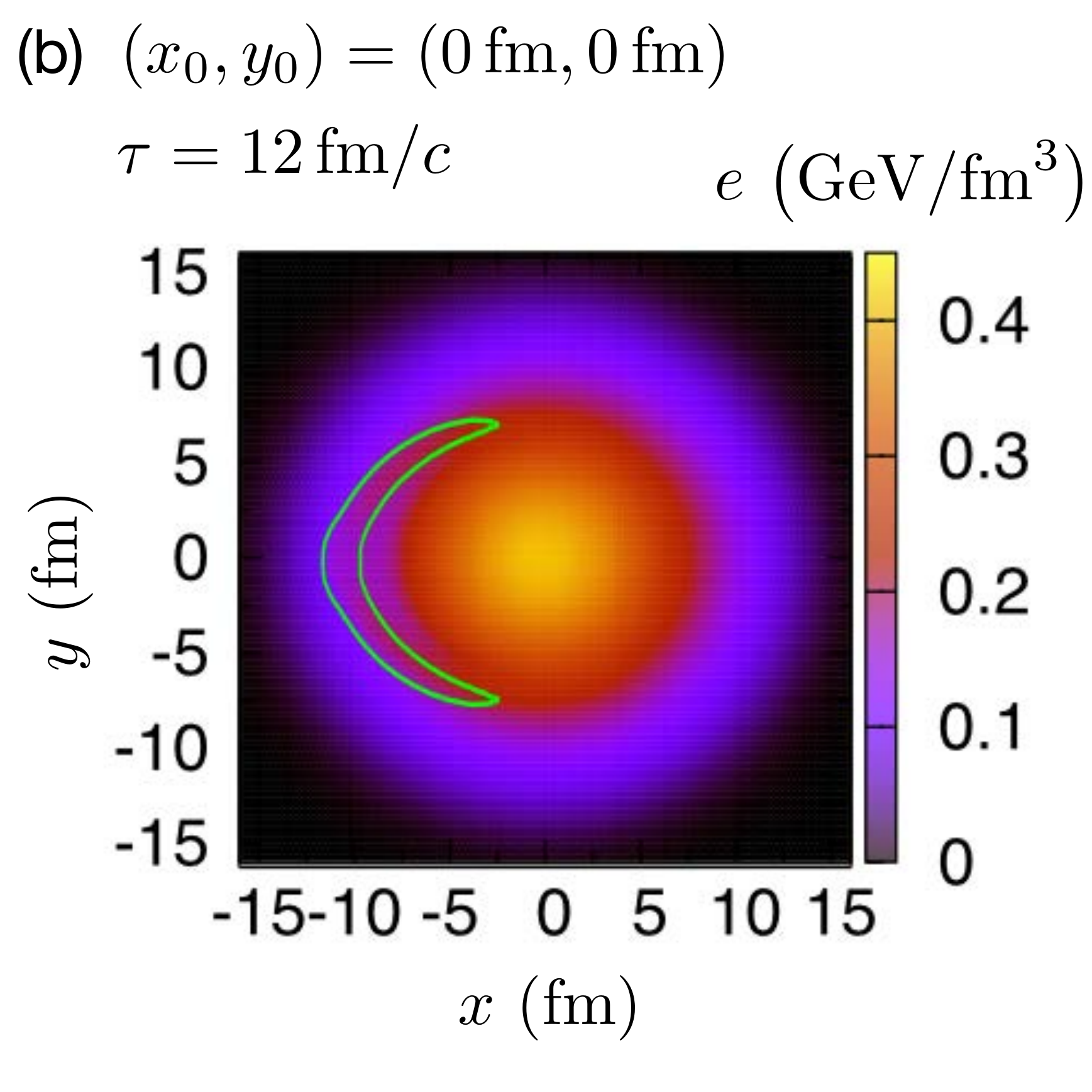}
  \hspace{14pt}
  \vspace{4pt}
  \includegraphics[width=5.4cm,bb=0 0 500 500]{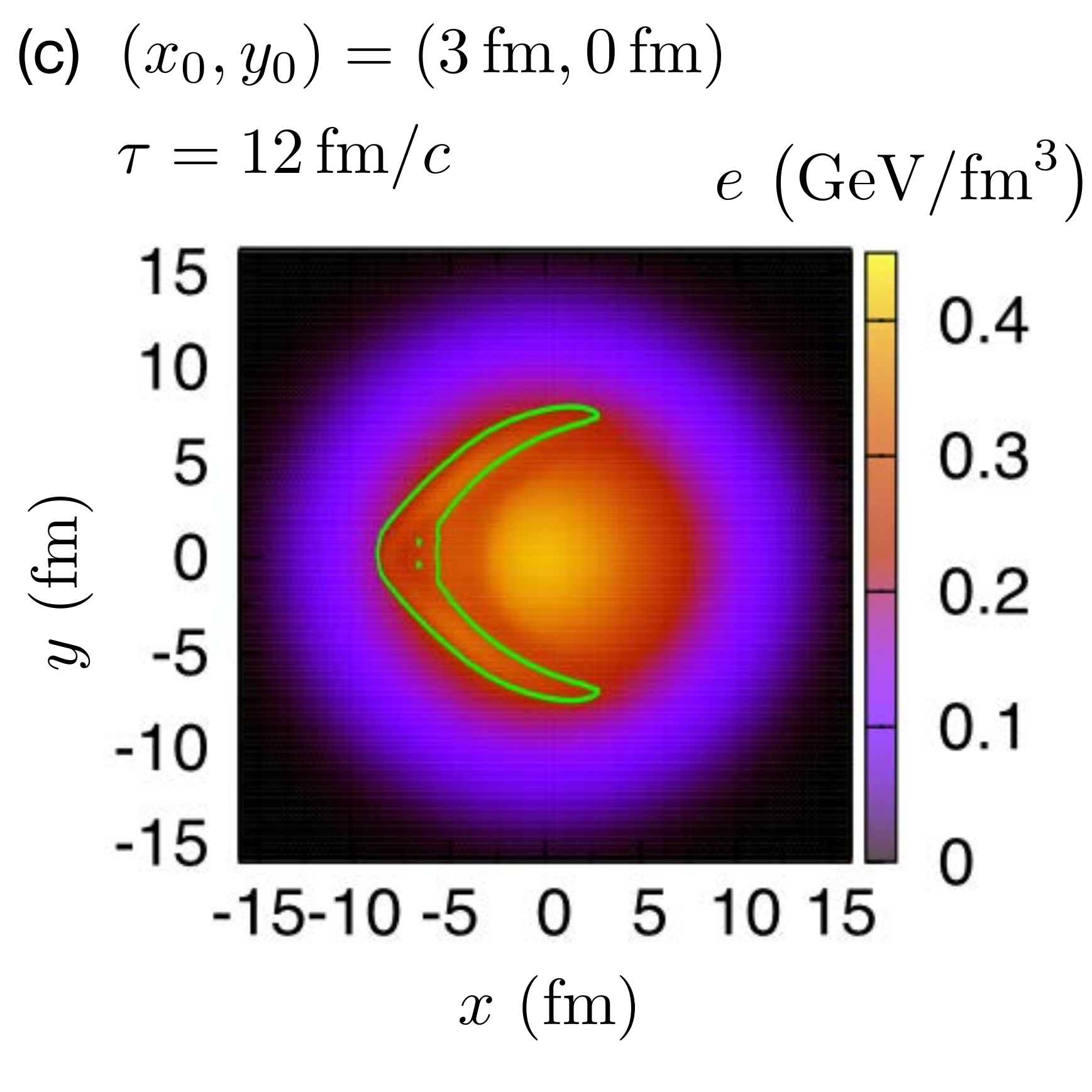}
     \hspace{14pt}
  \vspace{4pt}
   \includegraphics[width=5.4cm,bb=0 0 500 500]{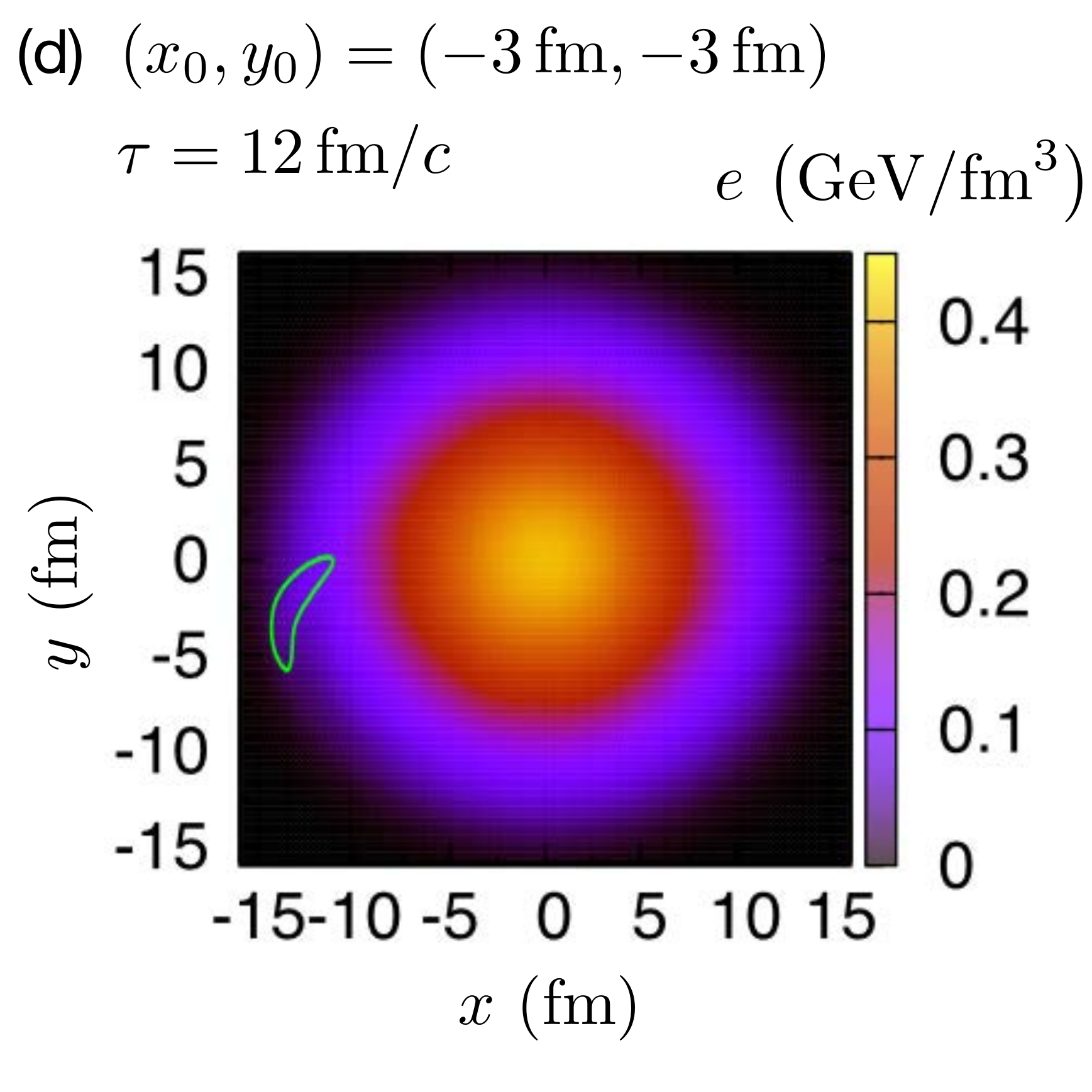}
  \hspace{14pt}
  \vspace{4pt}
   \includegraphics[width=5.4cm,bb=0 0 500 500]{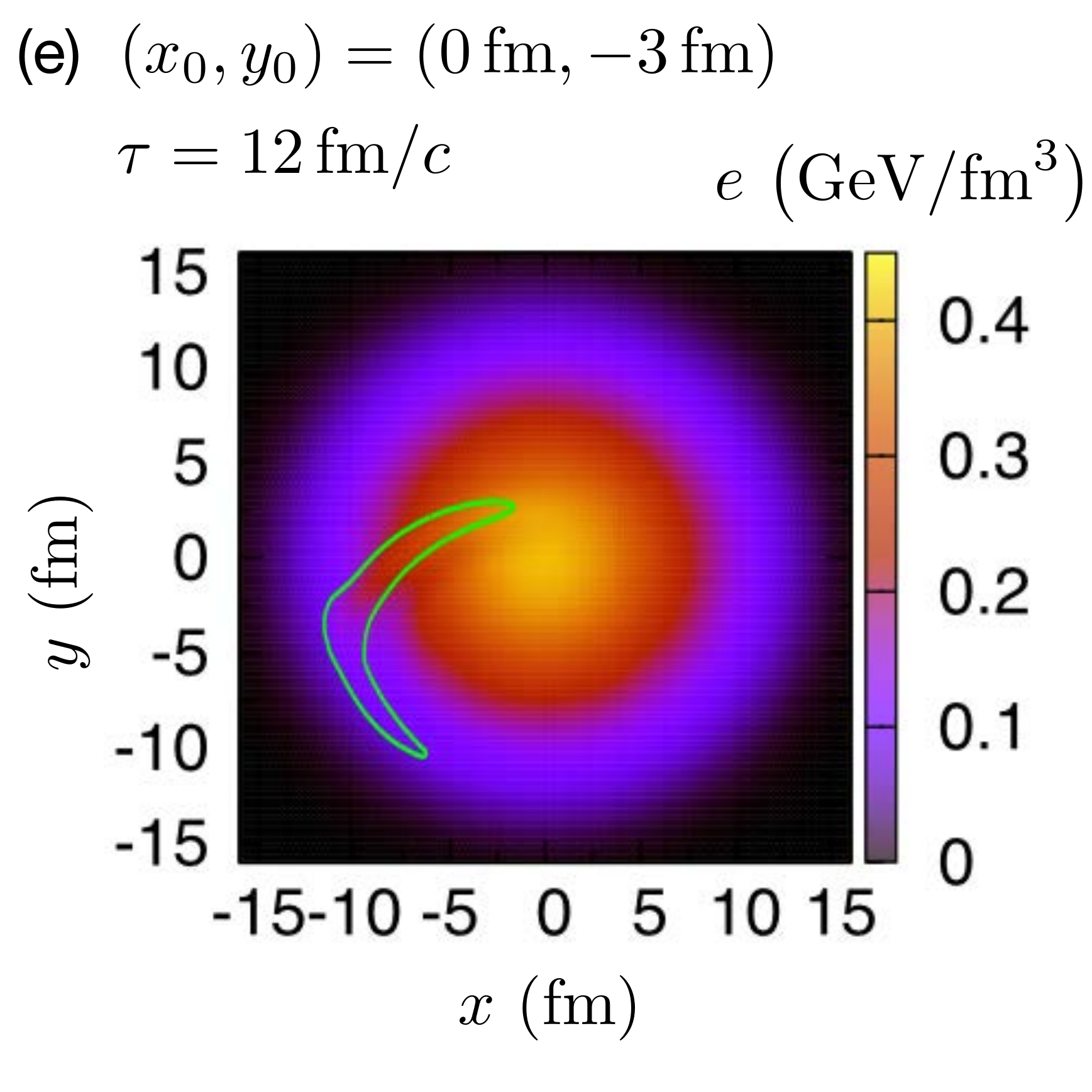}
  \hspace{14pt}
  \vspace{4pt}
   \includegraphics[width=5.4cm,bb=0 0 500 500]{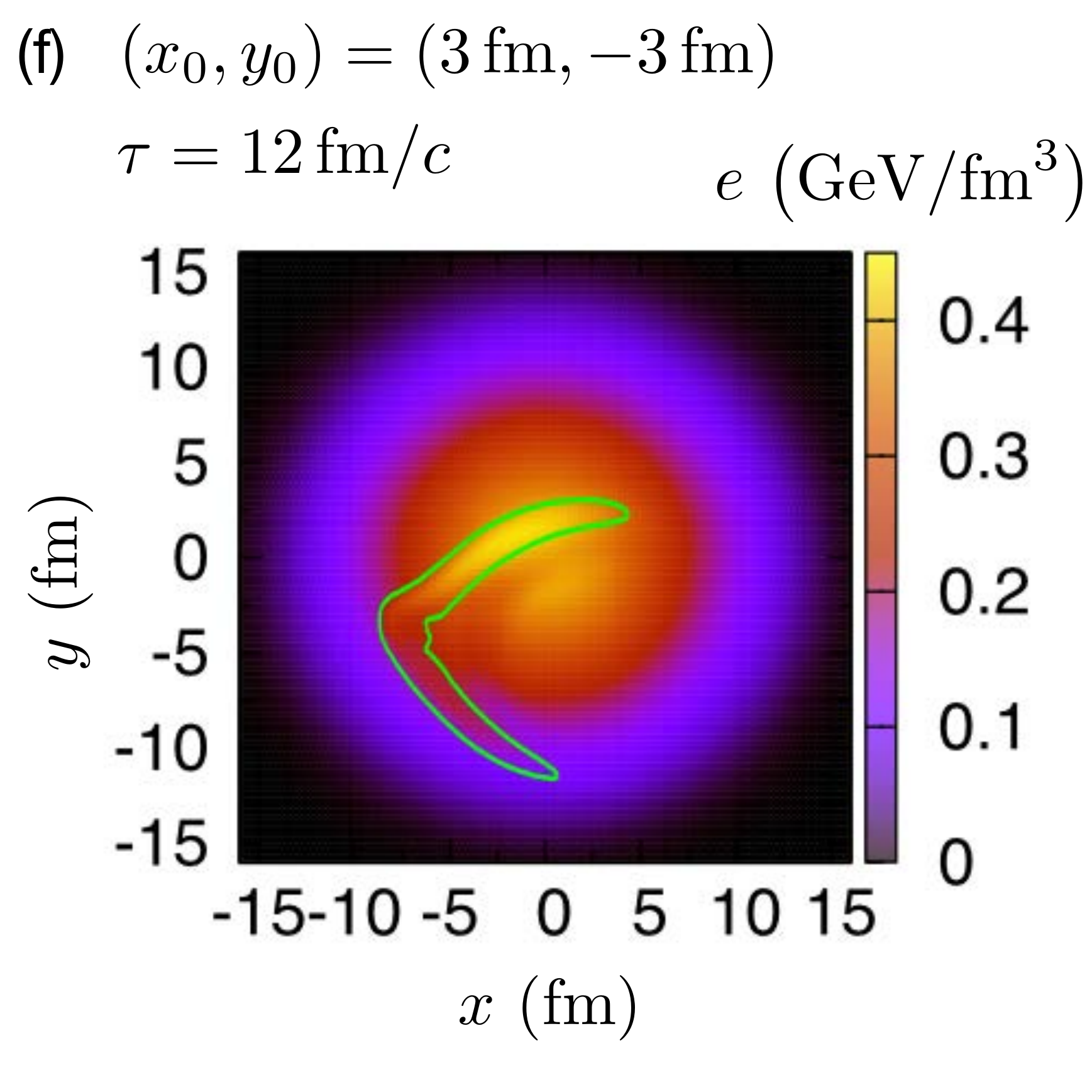}
  \hspace{14pt}
  \vspace{4pt}
   \includegraphics[width=5.4cm,bb=0 0 500 500]{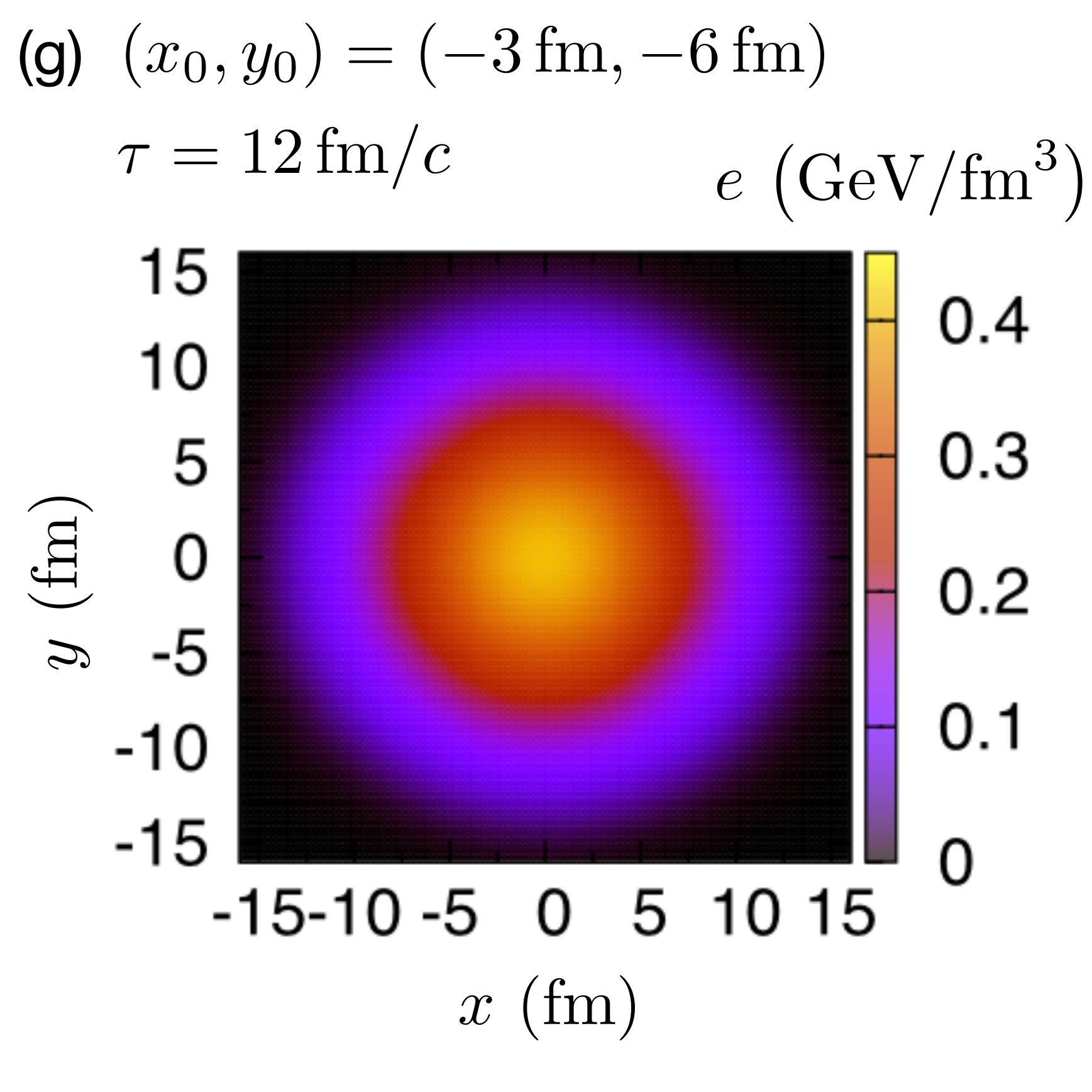}
  \hspace{14pt}
  \vspace{4pt}
   \includegraphics[width=5.4cm,bb=0 0 500 500]{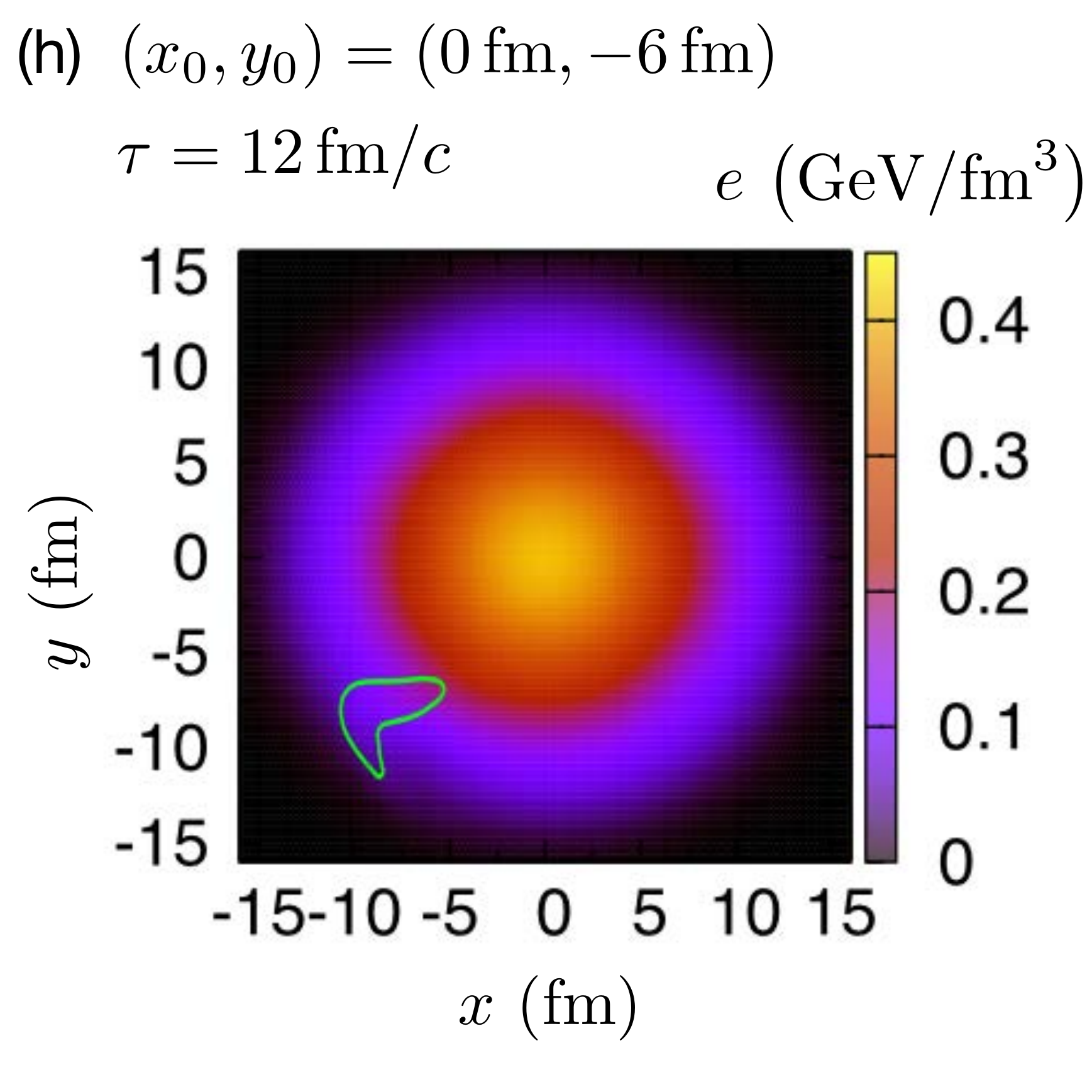}
  \hspace{14pt}
  \vspace{6pt}
   \includegraphics[width=5.2cm,bb=0 0 500 500]{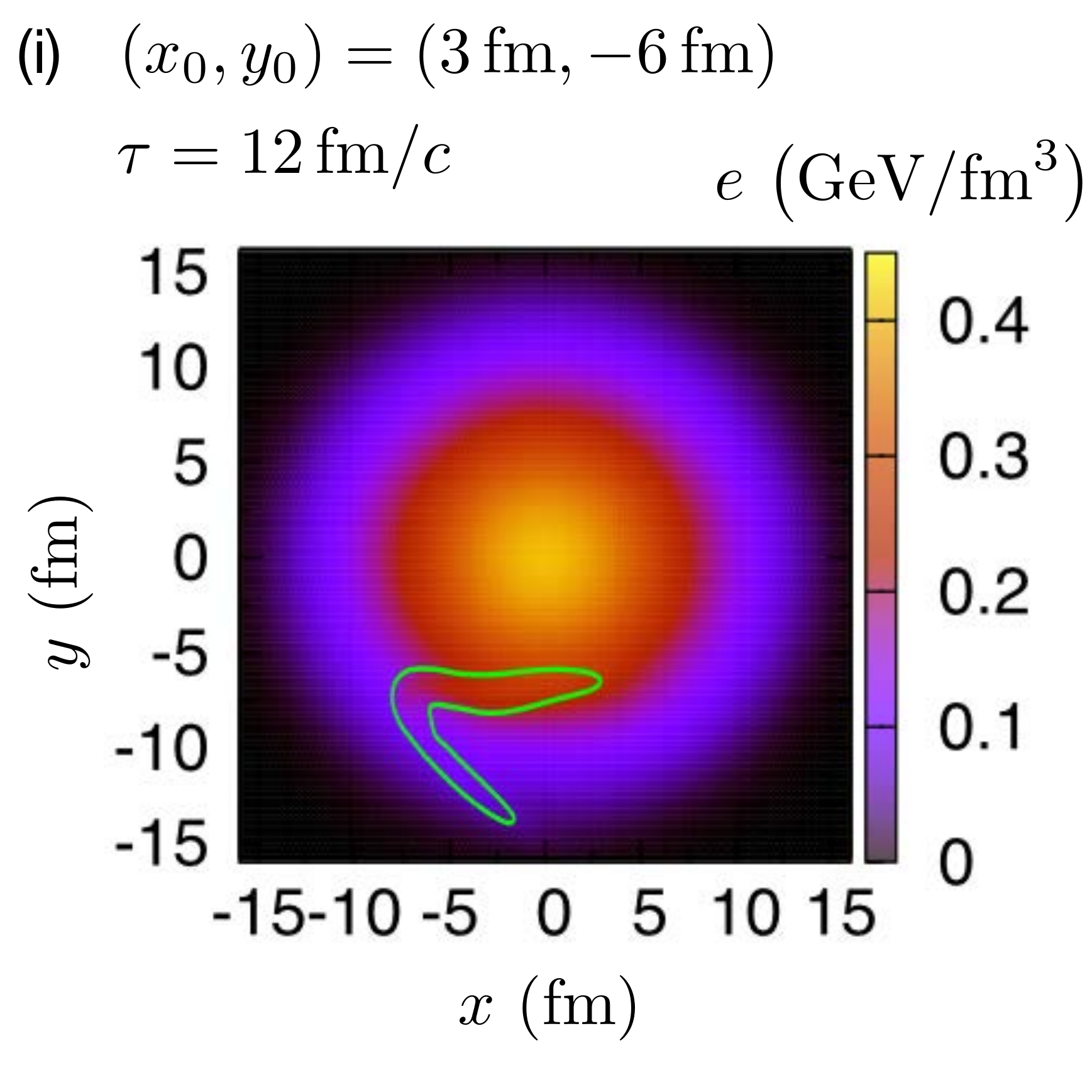}
  \caption{(Color online)
   Energy density distribution of the medium fluid 
   in the transverse plane at midrapidity $\str=0$ 
   for different jet production points. 
   The snapshots are taken at $\tau = 12.0\mbox{ fm}/c$. 
   The solid lines 
   show the higher energy density region 
   compared to the case without jet propagation. 
   }
    \label{fig:evo}
    \end{center}
 \end{figure*}

Figure \ref{fig:evo}
shows 
the energy density distributions 
of the medium fluid in the transverse plane at $\str=0$ 
at $\tau=12.0\,{\rm fm}/c$ 
for various production points for energetic partons. 
We can see relatively higher energy density regions 
which have U-shaped or V-shaped structures 
in Fig.~\ref{fig:evo} except for Fig.~\ref{fig:evo} (g). 
These are the remnants of Mach cones. 
The Mach cones 
are
induced by 
the energy momentum deposition from jets 
and 
distorted 
by the radial flow 
in various ways 
depending on 
where they develop 
in the medium. 
In the case where 
the jet is produced 
on the negative $x$ side 
and escapes from the radial flow, 
the radial flow 
pushes 
the wave front of the Mach cone 
mainly from the inside. 
Consequently, 
the shape of the Mach cone 
becomes rounded 
[Figs.~\ref{fig:evo} (a), (b), (d), and (e)]. 
In the case where the energetic parton travels through 
the off-central path in the medium, 
the Mach cone 
is drifted by the radial flow of the background 
and inclined to the inside of the medium.  
Furthermore, 
the Mach cone 
and 
the radial flow 
push each other, 
and thereby 
the Mach cone 
is asymmetrically distorted 
[Figs.~\ref{fig:evo} (f), (h), and (i)]. 
In an event where 
the jet is produced at 
the edge of the medium 
and then 
travels outward, 
the Mach cone 
is almost not formed 
[Fig.~\ref{fig:evo} (g)]. 
This is because 
the jet escapes from 
the hot region of the medium 
before the Mach cone develops. 

\subsection{Azimuthal angle distribution}
The main motivation 
in this paper 
is to study 
how 
the hydrodynamic response to jet propagation 
appears in 
the resulting particle spectra. 
We calculate 
the azimuthal angle distribution of 
charged pions 
emitted from the medium. 
Here, 
the distribution in the case without jet propagation 
is subtracted 
as a background: 
\begin{eqnarray}
\Delta \frac{dN_{{\rm \pi}^\pm}}{d\phi_p d\eta}
 &=& \frac{dN_{{\rm \pi}^\pm}}{d\phi_p d\eta}-
\left. \frac{dN_{{\rm \pi}^\pm}}{d\phi_p d\eta}\right|_{\mbox{no jet}}. 
\end{eqnarray}
In this study, 
we focus only on the low momentum particles 
originating from the bulk medium 
and 
do not include 
the particles from jet fragmentation 
in the calculations. 
When we take an event average, 
each event is weighted by 
the jet production rate 
as a function of 
the jet creation point 
and 
initial jet transverse momentum. 
As the weight for 
the spatial distribution of 
the jet production points 
in the transverse plane, 
we use 
the number density of the binary collisions 
between the nucleons in a Pb-Pb collision 
calculated from the Glauber model. 
As the momentum dependence of 
the jet production rate, 
we use a power function: 
\begin{eqnarray}
\frac{d^2 \sigma}{dp_{T,\rm jet} dy}&=&
\frac{1}{p_0}
\left(\frac{p_0}{p_{T, \rm jet}}\right)^{\alpha}, \label{eq:jetmomdis}
\end{eqnarray}
where $p_0=205 \,{\rm GeV/}c$ and $\alpha=6.43$  are parameters 
chosen to fit 
the data 
from the ATLAS Collaboration in $p$-$p$ collisions 
at $\sqrt{s_{NN}}=2.76\,{\rm TeV}$ 
\cite{Aad:2014bxa}. 

   \begin{figure}
      \begin{center}
 \includegraphics[width=8.0cm,bb=0 0 1024 768]{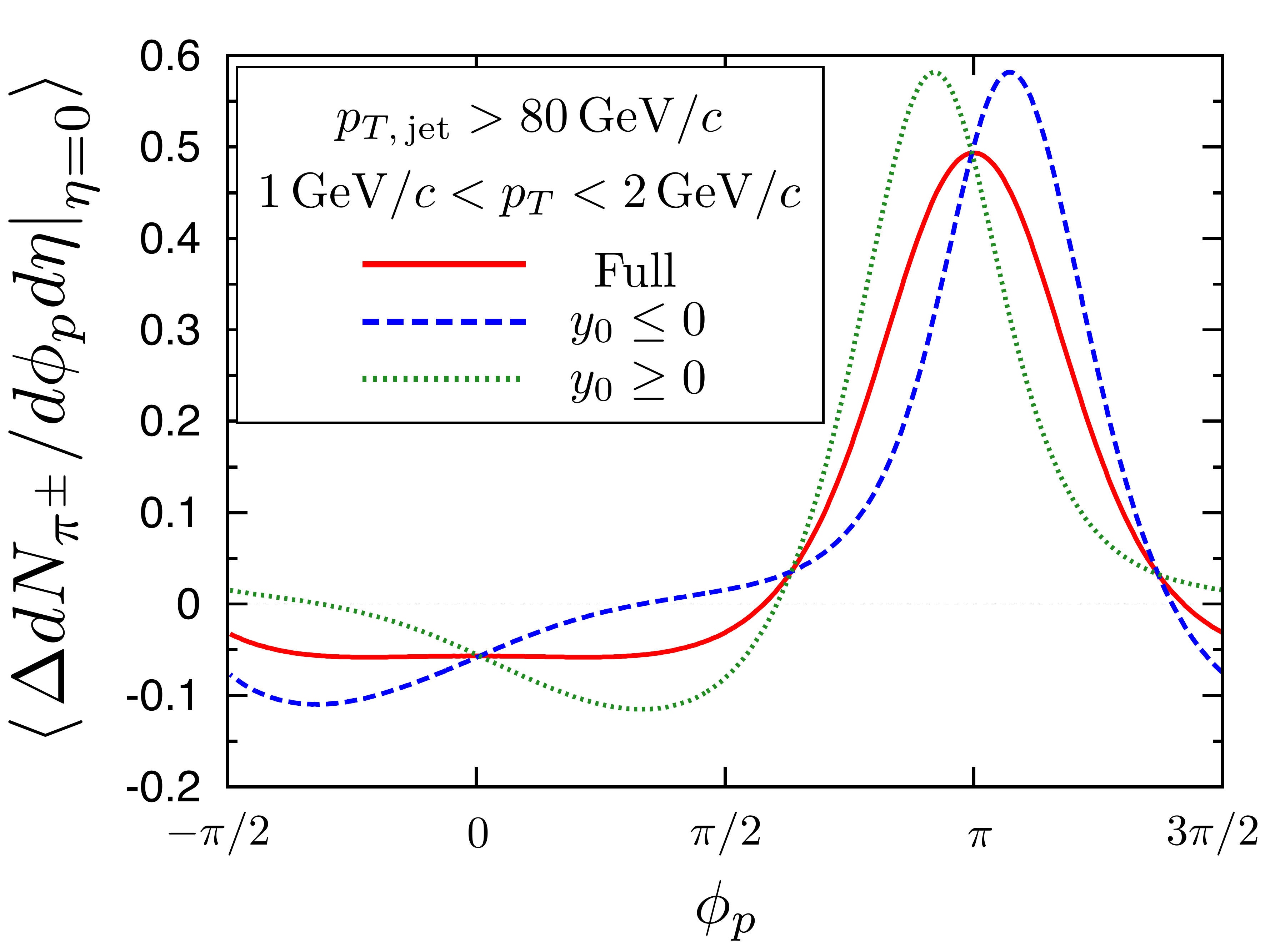}
    \caption{(Color online)
    Event-averaged
    azimuthal angle distributions 
    of charged pions with $1<p_{T}<2\,{\rm GeV}/c$ 
    subtracted by the background. 
    The trigger threshold for 
    the jet transverse momentum at final state 
    is 
    $p_{T\,\rm jet}>80\,{\rm GeV}/c$. 
    The solid line is
    the result 
    averaged over 
    all the triggered events. 
    The dashed line and the dotted line are
    the results 
    averaged over 
    events 
    with 
    the jet production points 
    in the region 
    $y\leq 0$ 
    and 
    in the region 
    $y\geq 0$, 
    respectively. 
    }
    \label{fig:event averaged}
          \end{center}
  \end{figure}
  
The solid line 
in Fig.~\ref{fig:event averaged} shows 
the azimuthal angle distribution 
at midrapidity 
after taking the event average. 
In this analysis, 
only the contribution of 
the charged pions 
with the transverse momenta 
between $1$ and $2$ GeV$/c$ 
is taken into account. 
This $p_T$ range is employed 
to measure the low-$p_T$ particles 
associated with large-$p_T$ jets 
in experiments by the CMS Collaboration \cite{Khachatryan:2016erx}. 
The average is taken over the events 
to the jet transverse momentum 
$p_{T,\,{\rm jet}}\geq 80\,{\rm GeV}/c$ 
in the final state.
We can see 
only a peak in the jet direction $\phi_p=\pi/2$, 
and 
the opposite side around $\phi_p=0$ 
is almost flat. 
Intuitively, 
Mach cones are expected 
to produce a double-hump structure 
around the jet direction 
in the azimuthal distribution 
by an analogy to a ring image of 
the Cherenkov radiation. 
However, such a structure 
cannot be obtained 
from 
this hydrodynamic calculation. 
We have some reasons for 
the absence of the double-hump structure. 
First, 
the momentum deposition of the jet 
induces 
the diffusion wake 
which 
produces 
a prominent peak 
in the spectra in the direction of jets. 
Second, 
when calculating 
the spectra 
through the Cooper-Frye formula (\ref{eqn:C-F}), 
the thermal equilibrium distribution 
boosted by the flow velocity 
is used. 
The flow velocity distribution 
is not directly reflected 
and 
is smeared out 
in the consequent spectra.
Third, 
since the Mach cones are largely distorted 
by the expansion of the medium, 
the Mach cones 
no longer have 
the clear conical shock front. 
Actually, 
the double-hump structure 
caused by the Mach cone 
does not exist in the observation 
by the CMS Collaboration 
\cite{Khachatryan:2016erx}. 
In this result, any clear signal reflecting the characteristic structure of a Mach cone cannot be seen. 

The dashed line and the dotted line 
in Fig.~\ref{fig:event averaged} show 
the results 
averaged over 
events 
where 
the jet production points 
are restricted 
in the region 
$y\leq 0$ 
and 
in the region 
$y\geq 0$, 
respectively. 
These are 
symmetric with respect to 
$\phi=0$ and $\pi$ 
due to the symmetry 
of the geometry 
across the $z$-$x$ plane. 
The average of them 
is equal to 
the result 
averaged over 
the full events. 
The peaks 
are shifted 
to the direction of 
the radial expansion 
and, furthermore, 
dips 
can be seen 
in specific directions 
almost perpendicular to the peaks. 
The appearance of such structures 
in the azimuthal angle distribution 
reflects 
the geometrical relation 
between the jet production point 
and the medium. 

   \begin{figure}
       \begin{center}
    \includegraphics[width=8.0cm,bb=0 0 1024 768]{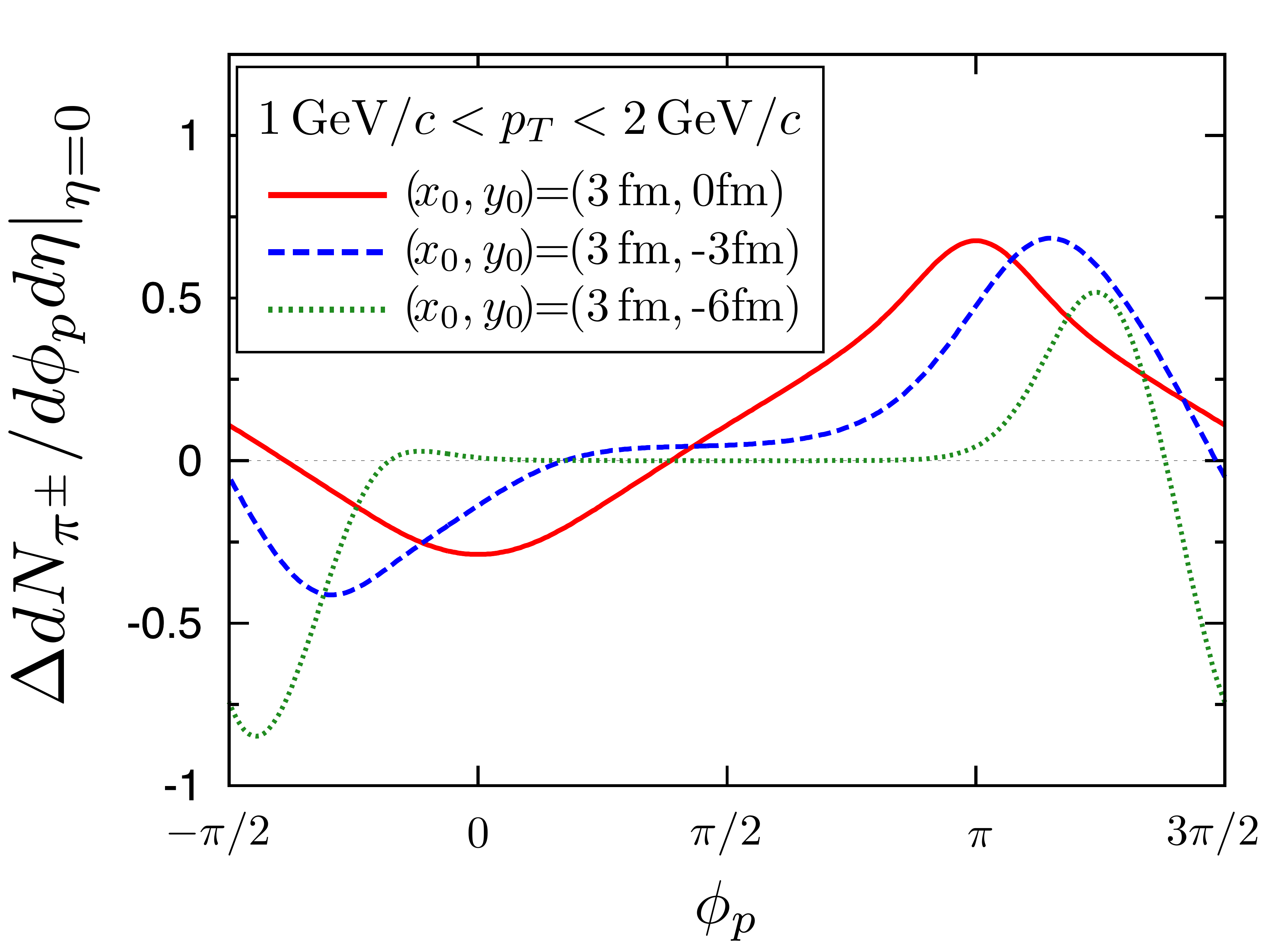} 
    \caption{(Color online)
    Azimuthal angle distributions 
    of charged pions with $1<p_{T}<2\,{\rm GeV}/c$ 
    subtracted by the background. 
    The solid line, 
    the dashed line, and
   the dotted line are 
   the results for the events with 
  the jet production point at 
  $\left(x_0,y_0\right)=\left(3.0\,{\rm fm},0\,{\rm fm}\right)$, 
  $\left(3.0\,{\rm fm},-3.0\,{\rm fm}\right)$, and 
  $\left(3.0\,{\rm fm},-6.0\,{\rm fm}\right)$, 
   respectively. 
    }
    \label{fig:x3}
            \end{center}
  \end{figure}
  
  \begin{figure}
      \begin{center}
    \includegraphics[width=8.0cm,bb=0 0 1024 768]{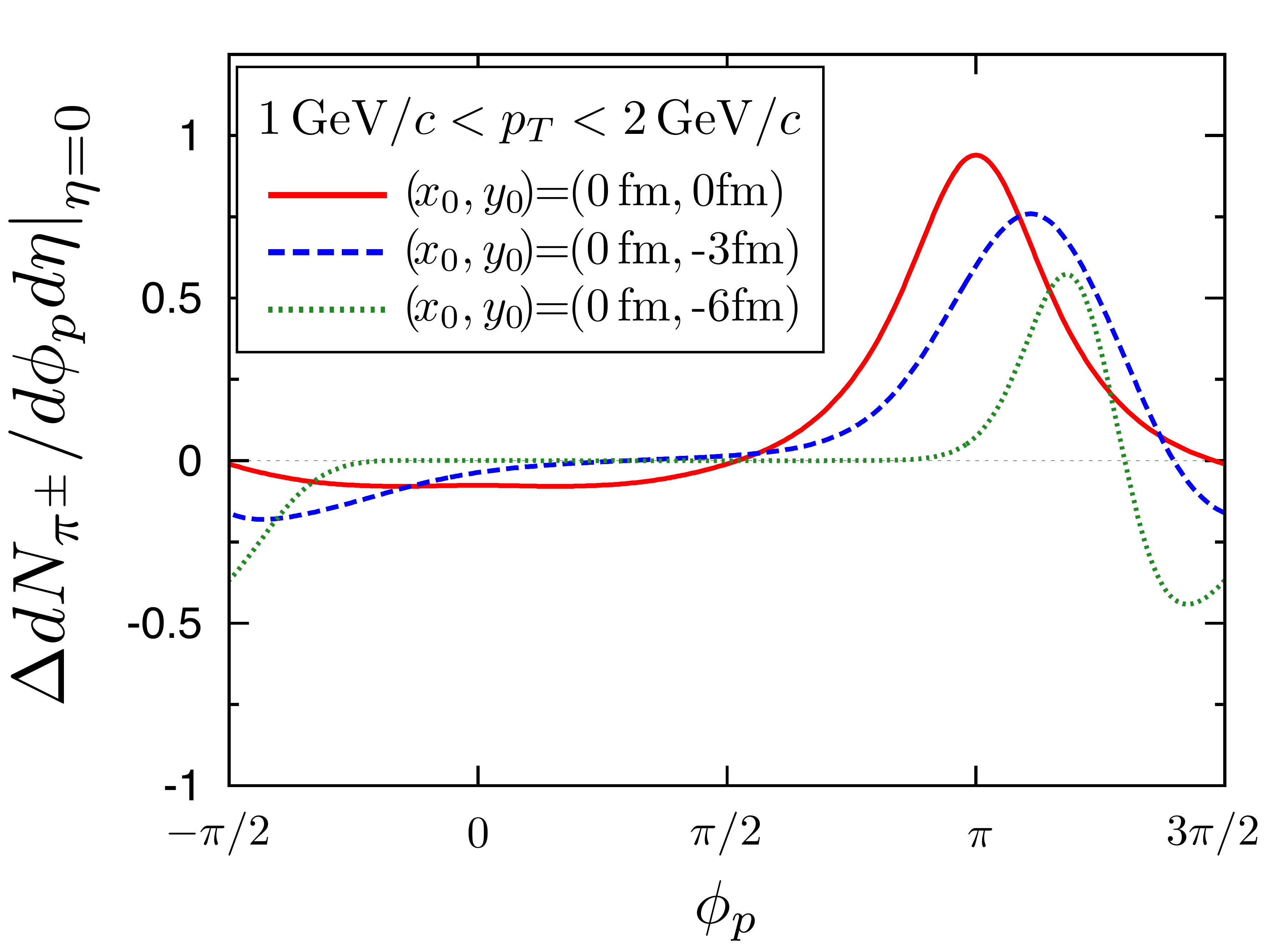} 
   \caption{(Color online)
    Azimuthal angle distributions 
    of charged pions with $1<p_{T}<2\,{\rm GeV}/c$ 
    subtracted by the background. 
    The solid line, 
    the dashed line, and
   the dotted line are 
   the results for the events with 
  the jet production point at 
  $\left(x_0,y_0\right)=\left(0\,{\rm fm},0\,{\rm fm}\right)$, 
  $\left(0\,{\rm fm},-3.0\,{\rm fm}\right)$, and 
  $\left(0\,{\rm fm},-6.0\,{\rm fm}\right)$, 
   respectively. 
   }
    \label{fig:x0}
            \end{center}
  \end{figure}
  
  \begin{figure}
    \begin{center}
   \includegraphics[width=8.0cm,bb=0 0 1024 768]{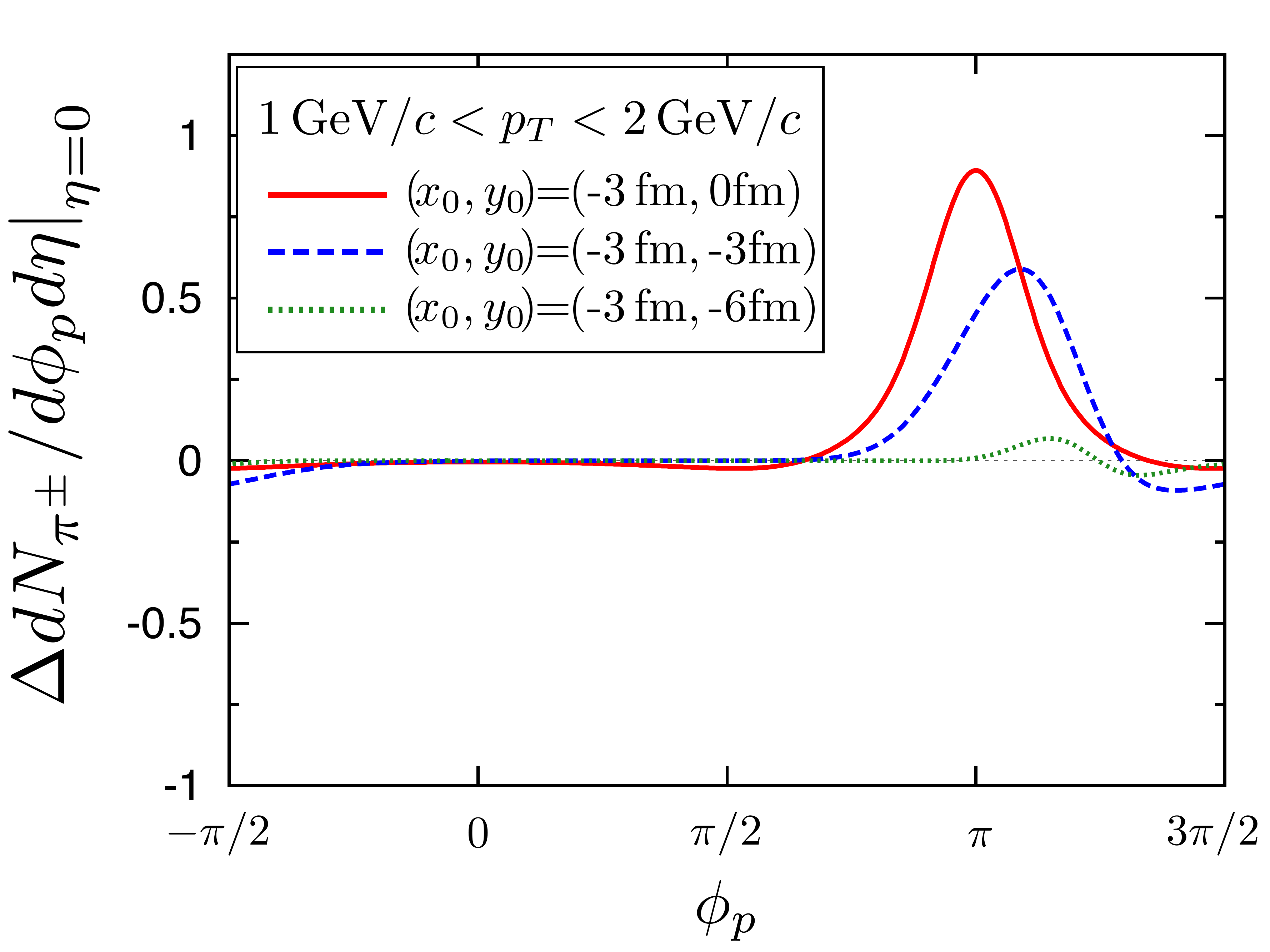} 
   \caption{(Color online)
    Azimuthal angle distributions 
    of charged pions with $1<p_{T}<2\,{\rm GeV}/c$ 
    subtracted by the background. 
    The solid line, 
    the dashed line, and
   the dotted line are 
   the results for the events with 
  the jet production point at 
  $\left(x_0,y_0\right)=\left(-3.0\,{\rm fm},0\,{\rm fm}\right)$, 
  $\left(-3.0\,{\rm fm},-3.0\,{\rm fm}\right)$, and 
  $\left(-3.0\,{\rm fm},-6.0\,{\rm fm}\right)$, 
   respectively. 
   }
    \label{fig:x-3}
        \end{center}
  \end{figure}
  
To interpret these structures, 
let us consider specific events. 
Figures \ref{fig:x3}-\ref{fig:x-3} 
show 
the azimuthal angle distributions 
for a single event 
with a specific jet production point. 
Figure \ref{fig:x3} 
shows the results 
in the case 
where the jet production point 
is 
located in the fourth quadrant 
$(x_0=3\,{\rm fm},y_0=0,-3,-6\,{\rm fm})$. 
We can see that the peak 
in the jet direction 
shifts 
to larger $\phi$ 
as the distance 
between 
the jet production point 
and 
the $x$ axis 
increases. 
This is because 
the fast flow 
of the diffusion wake 
is pushed and bent 
by 
the radial flow 
of the background medium. 
At the edge of the medium, 
the radial flow 
becomes faster 
and 
pushes the diffusion wake 
strongly. 
When the jet is produced 
at $(x_0,y_0)=(3\,{\rm fm},0\,{\rm fm})$, 
the number of 
low $p_T$ particles in the direction opposite to the jet
($\gamma$ direction) 
decreases. 
While the jet travels toward 
the center of the medium, 
the Mach cone 
propagates
squarely against 
the radial flow. 
As a result, 
the particle emission 
in the $\gamma$ direction 
is suppressed. 
In Fig.~\ref{fig:x3}, 
one can see 
a dip at $\phi_p\sim-\pi/4$ 
in the case of the jet produced 
at $(x_0,y_0)=(3\,{\rm fm},-3\,{\rm fm})$ and 
at $\phi_p\sim-\pi/2$ 
in the case of 
$(x_0,y_0)=(3\,{\rm fm},-6\,{\rm fm})$. 

  \begin{figure}
    \begin{center}
    \vspace{10pt}
    \hspace{-25pt}
   \includegraphics[width=7.5cm,bb=0 0 1024 768]{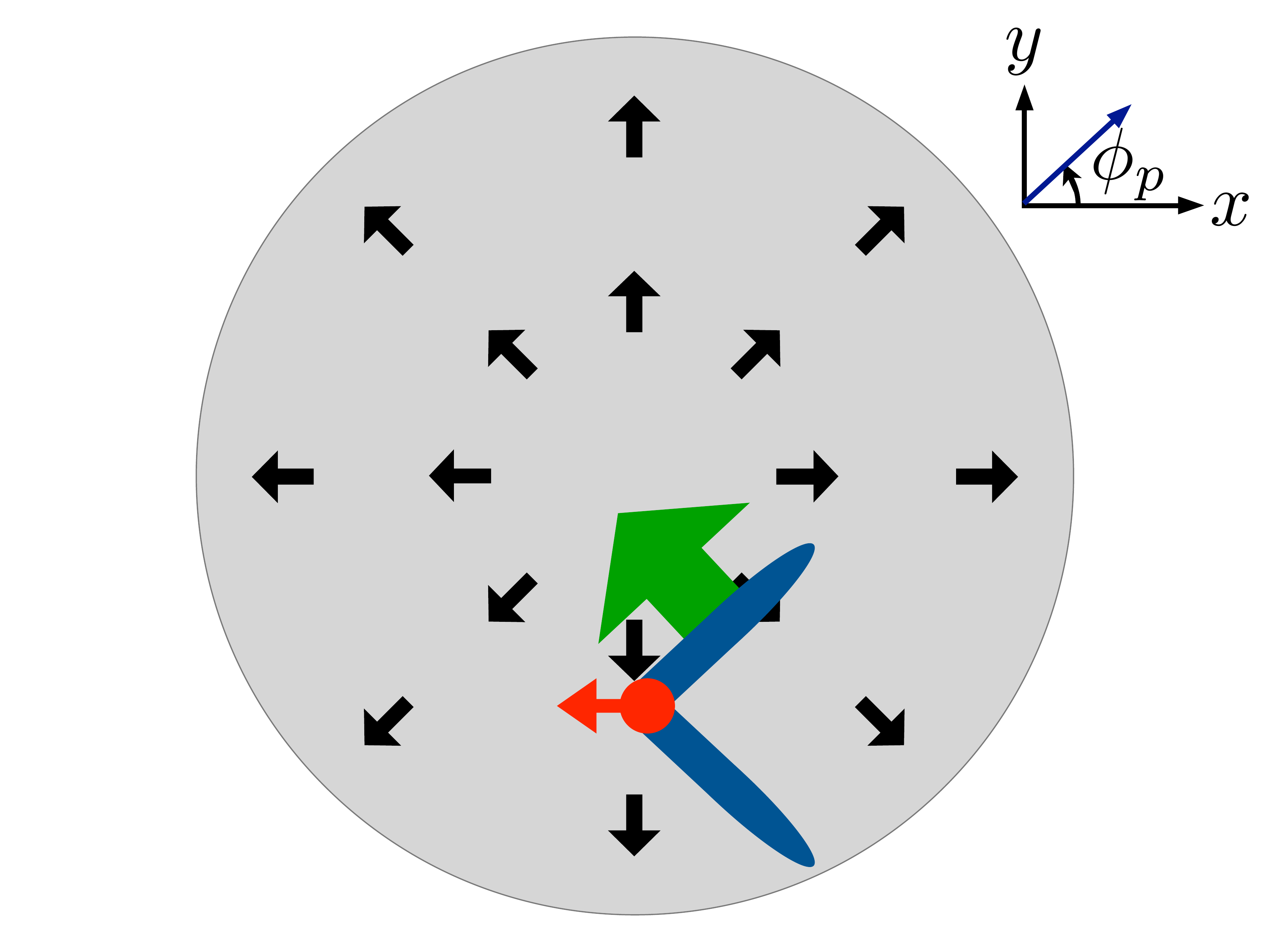} 
    \caption{(Color online)
      Schematic picture of the transverse plane $\eta_{\rm s}=0$ 
      to illustrate 
      how the Mach cone propagating 
      in the expanding QGP suppresses the particles 
      in a certain direction. 
      The radial flow is induced by the expansion of the background QGP medium 
      (arrows pointing in the radial direction). 
      The jet (small circle with a small arrow at the top of the V-shaped region) travels through the lower half region of the QGP 
      and induces the Mach cone (V-shaped region). 
      During propagation, the wave front of the Mach cone 
      pushes back the radial flow 
      (large arrow from the side of the V-shaped region). 
      As a result, the particles emitted 
      in the direction opposite to the large arrow 
      are suppressed. 
    }
    \label{fig:manga}
  \end{center}
    \end{figure}
    
Figure \ref{fig:manga} 
is an illustration 
of how this dip 
appears 
in a certain direction 
in the azimuthal angle distribution. 
When 
the jet path is away from 
the center of the medium, 
the induced Mach cone 
violates the symmetry 
of the flow profile
with respect to the $x$ axis 
(V-shaped region in Fig.~\ref{fig:manga}). 
In particular, 
the wave front of the Mach cone 
on the center side of the medium 
develops 
more largely 
than that on the outer side 
because of the higher temperature 
and 
pushes back the radial flow strongly 
in a certain direction 
(large arrow from the side of the V-shaped region in Fig.~\ref{fig:manga}). 
Thereby, 
in the direction in which 
the radial flow 
is held back 
by the wave front on the center side, 
the particle emission 
is suppressed. 

Figure \ref{fig:x0} 
shows the results 
when the jet production point 
is set on the negative $y$ axis 
$(x_0=0\,{\rm fm},y_0=0,-3,-6\,{\rm fm})$. 
The shifts of the peaks 
in the jet direction 
and 
the dips 
can also be seen. 
The angle between 
the dip direction and the $\gamma$ direction
is large 
compared to the ones 
in the case 
of the jet production point 
in the fourth quadrant. 
This is because 
the direction 
of the radial flow 
pushed back by 
the Mach cone 
is 
shifted 
to the clockwise direction 
depending on the jet path in the medium. 
In Fig.~\ref{fig:x-3}, 
the results 
in the case 
of the jet produced 
in the third quadrant 
$(x_0=-3\,{\rm fm},y_0=0,-3,-6\,{\rm fm})$ 
are shown. 
The structures 
in the $\gamma$ direction 
are almost flat. 
In these cases, 
the jet propagates away from
the radial flow. 
The induced Mach cone 
is pushed 
mainly from the inside 
by the radial flow 
in the jet direction 
and 
does not affect 
the radial flow 
in the $\gamma$ direction. 

As mentioned above, 
the interplay 
between 
Mach cone and radial expansion 
appears as a dip in azimuthal distributions. 
It 
can be 
referred to 
as 
the signal of 
the Mach cone 
and 
its 
direction and depth 
vary with 
the jet path 
in the expanding QGP medium. 
The trigger threshold for jets 
can constrain 
the jet production point 
\cite{Dainese:2004te,Renk:2006nd,Renk:2006pk, Qin:2012gp}. 
Jets with lower energies 
are produced much more frequently 
because 
the production rate 
is a steeply decreasing function of the jet energy. 
However, 
when jets travel through the center of the medium, 
jets with low initial energy 
are hardly triggered 
in the final state 
because of the large energy loss. 
Therefore, 
if the trigger 
is set 
to 
choose 
small-energy-loss events, 
the distribution of the jet production point 
in the triggered events 
is 
biased 
to 
the surface of the medium. 
We here show 
how 
the distribution of the jet production point 
and 
the resulting 
azimuthal angle distribution 
of the soft particles 
are 
varied 
depending on the trigger threshold. 

\begin{figure}
  \begin{center}
  \includegraphics[width=8.0cm,bb=0 0 1024 768]{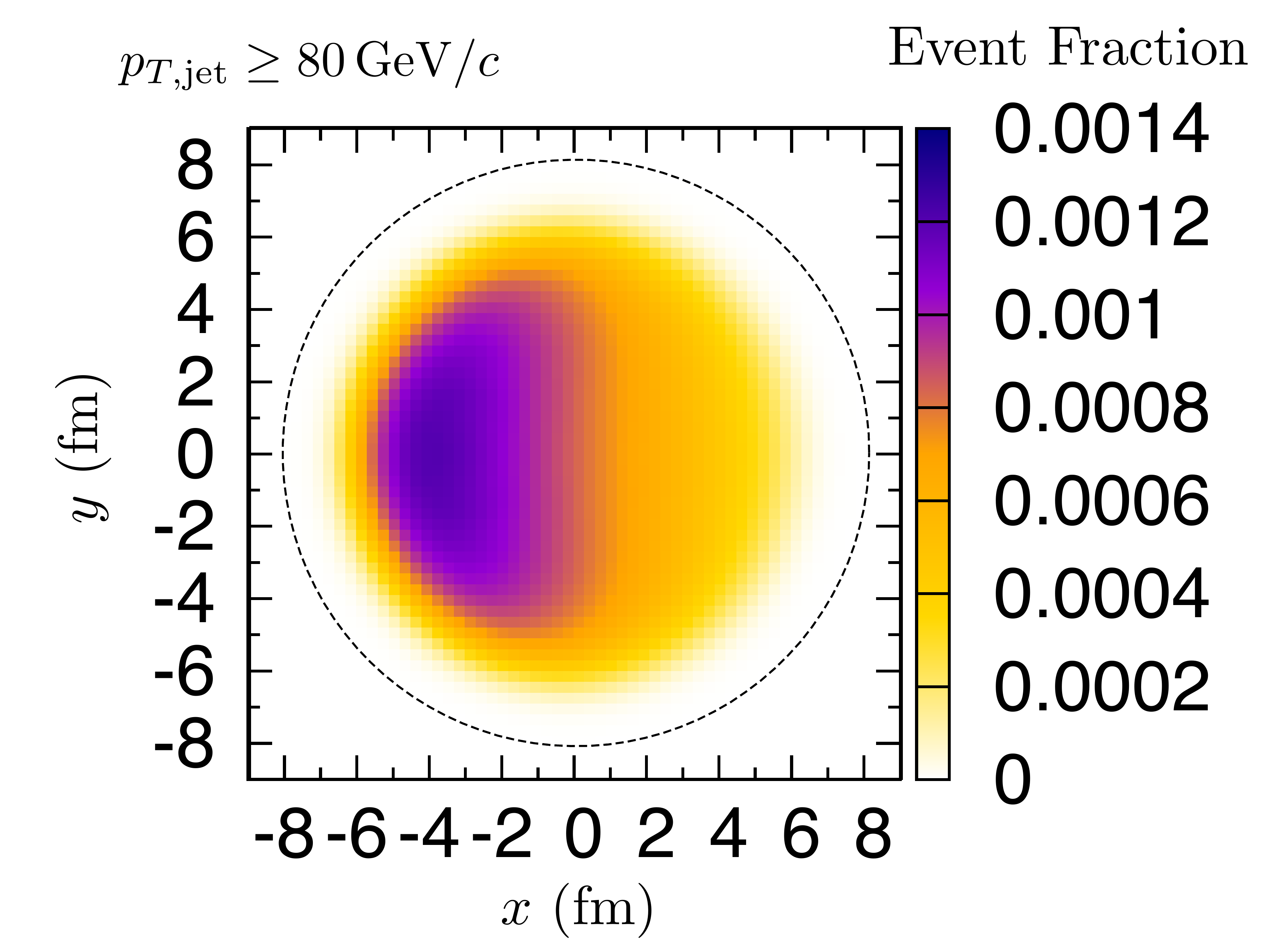}
   \caption{(Color online)
    Distribution of the jet production point 
    for the $\gamma$-jet events 
    when the trigger threshold 
    for the jet transverse momentum 
    in the final state 
    is set to $80\,{\rm GeV}/c$. 
   }
    \label{fig:map1}
    \end{center}
  \end{figure}

Figure \ref{fig:map1} 
shows 
the distribution of the jet production point 
when the trigger threshold 
for the jet transverse momentum 
in the final state 
is set to 
$p_{T,\,\rm jet}>80\,{\rm GeV}/c$. 
The jet particles are produced likely 
near the central region 
because 
the number of binary collisions 
between the nucleons 
is a maximum at the center 
and goes to zero at the edge 
of the reaction region. 
However, 
we can see that 
the events 
in which 
the jet is produced 
near the surface of the medium 
and then escapes to the outside 
are dominant. 
This is because 
the lower energy jets 
which are produced more frequently 
can survive 
with enough transverse momenta 
to be triggered 
in such events. 

\begin{figure*}
\begin{center}
   \includegraphics[width=8.0cm,bb=0 0 1024 850]{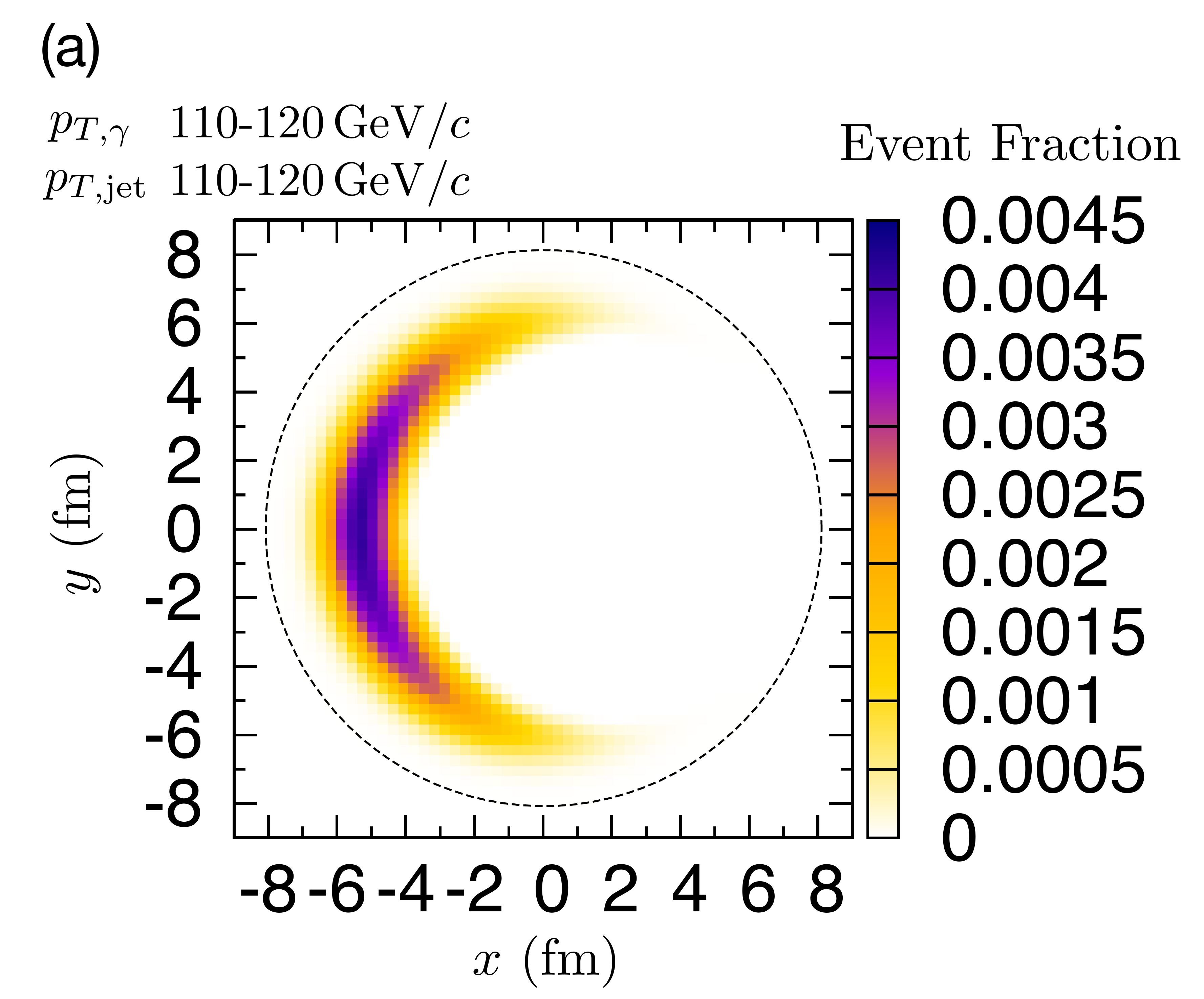}
  \hspace{16pt}
  \vspace{16pt}
      \includegraphics[width=8.0cm,bb=0 0 1024 850]{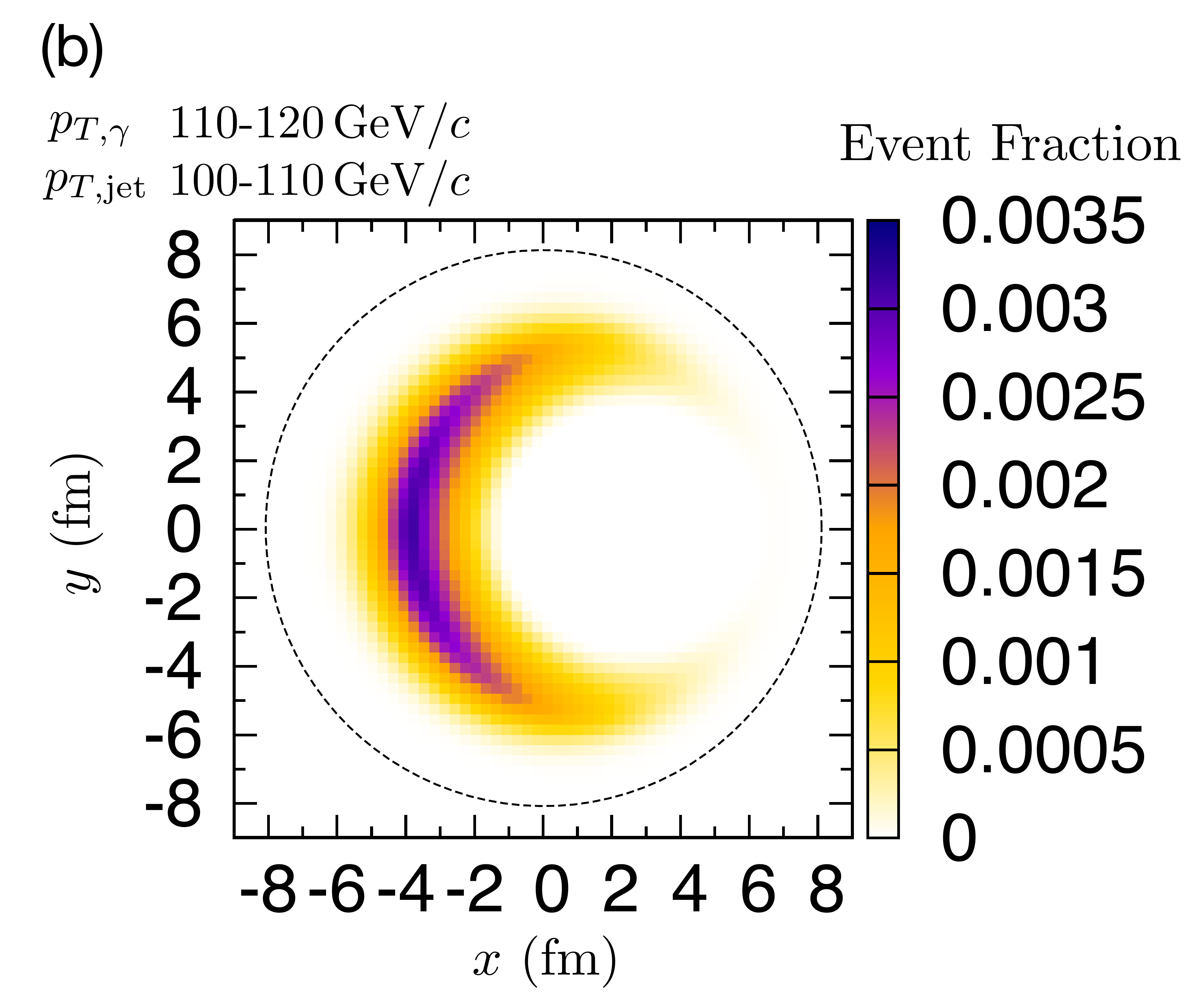}
  \hspace{16pt}
  \vspace{16pt}
   \includegraphics[width=8.0cm,bb=0 0 1024 850]{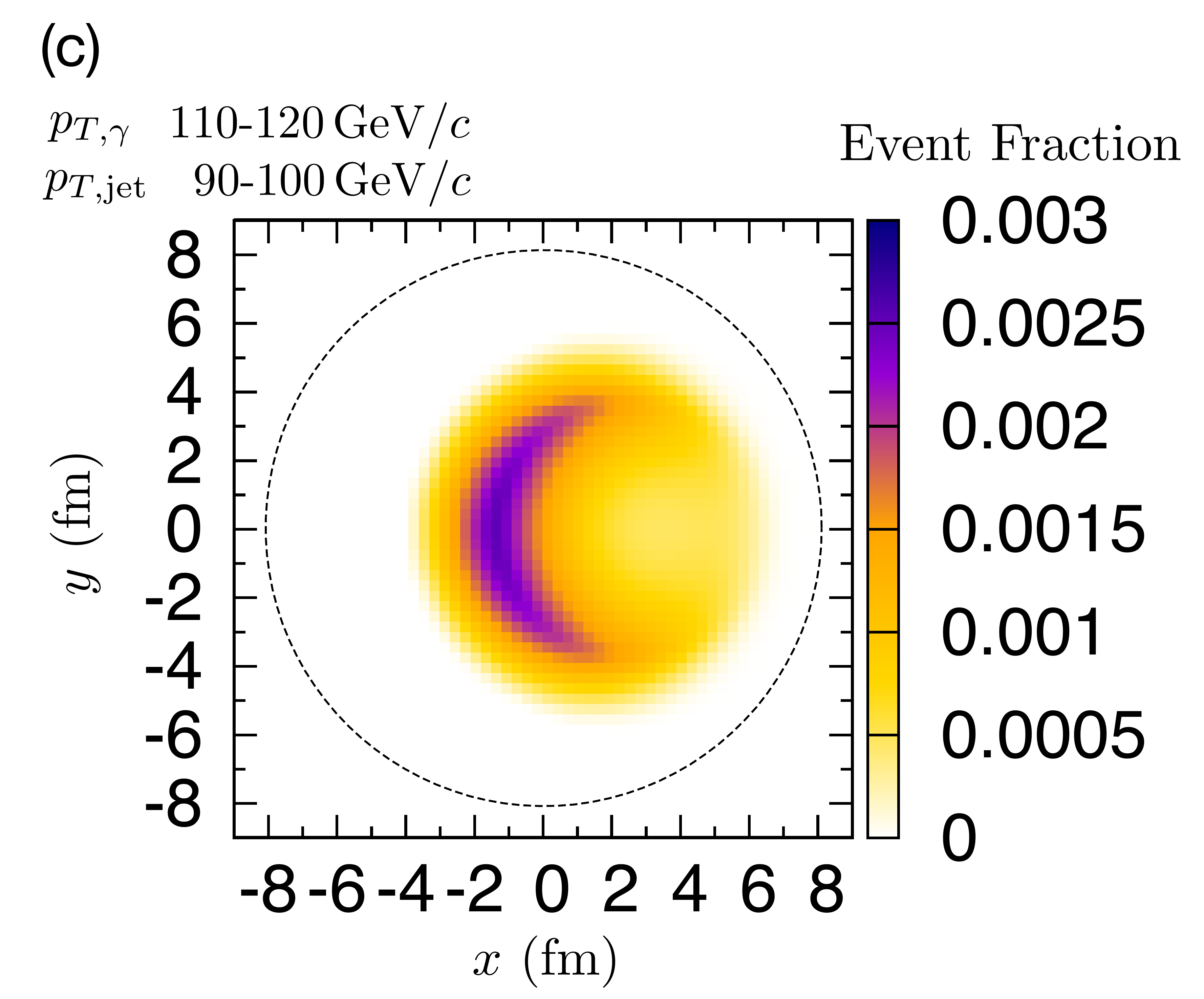}
  \hspace{16pt}
  \vspace{16pt}
  \includegraphics[width=8.0cm,bb=0 0 1024 850]{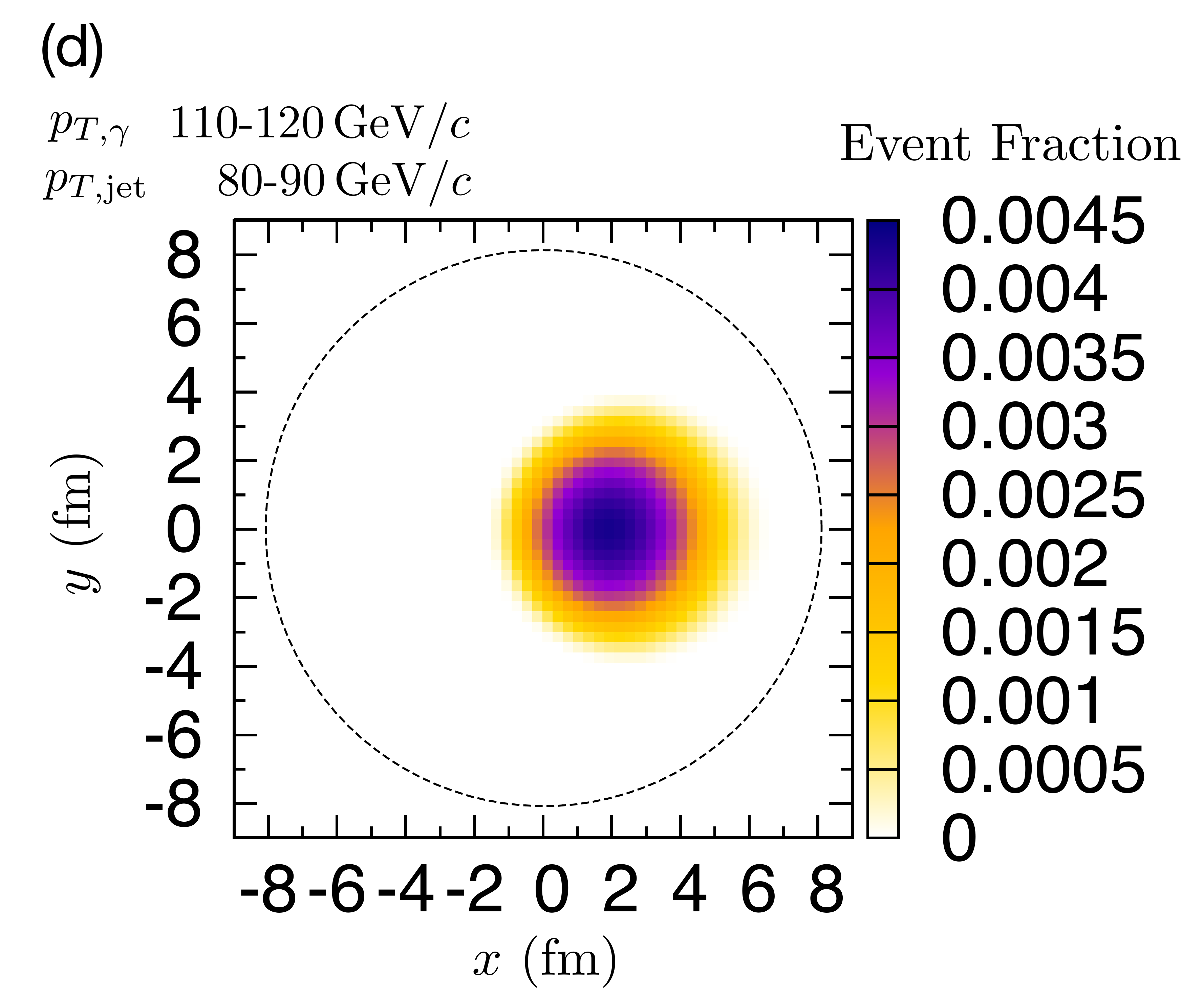}
   \caption{(Color online)
   Distribution of the jet production point 
   for different transverse momentum ranges 
   of the jet trigger. 
   The trigger range of the photon transverse momentum 
   is $110$-$120$ GeV$/c$. 
   The trigger ranges of the jet transverse momentum 
   are (a) $110$-$120$ GeV$/c$, 
   (b) $100$-$110$ GeV$/c$, 
   (c) $90$-$100$ GeV$/c$, and 
   (d) $80$-$90$ GeV$/c$. 
}
    \label{fig:map}
    \end{center}
 \end{figure*}

Next, we 
introduce 
trigger thresholds for 
the transverse momentum 
of the jet 
both in the final state 
and 
in the initial state. 
We can regard 
the initial transverse momentum 
of the jet 
as the transverse momentum 
of the photon 
$p_{T,\gamma}$ 
in a $\gamma$-jet event. 
Here, 
the trigger 
of the photon 
is fixed so that 
its transverse momentum 
is $100$-$120$ GeV$/c$. 
Figure \ref{fig:map} 
shows 
the distribution of the jet production point 
for different $p_{T,\,\rm jet}$ ranges 
of the trigger. 
When the trigger is set 
to extract 
small energy loss events, 
the detected jets are 
created 
in the crescent-shaped region 
near the surface of the medium 
[Figs.~\ref{fig:map} (a) and (b)]. 
However, when 
large-energy-loss events 
are extracted 
by the trigger, 
the detected jets are 
created in the region 
near the center 
in which 
the path length in the medium 
is to be long 
[Figs.~\ref{fig:map} (c) and (d)]. 
Ratios 
of the number of these events 
to that when the jet trigger is set to $p_{T,\rm jet} > 80$ GeV/$c$ 
are shown in Table \ref{tab}. 
The middle column is the ratio to the number of events
when the trigger also for photon transverse momentum is set to $110$-$120$ GeV/$c$.
The sum of these ratios is equal to unity. 
The right-hand column is the ratio to that without the trigger for the photon 
which corresponds to the jet production point distribution shown in Fig. \ref{fig:map1}. 

 \begin{table}
  \begin{center}
      \caption{
    Ratios of the number of events with triggers for both photon 
($110< p_{T,\gamma} <120$ GeV/$c$) and jet momenta shown in Figs. \ref{fig:map} (a)-(d)
to that (middle column) with the triggers 
        for both photon ($110< p_{T,\gamma} < 120$ GeV/$c$) and jet ($p_{T,\rm jet} > 80$ GeV/$c$)  and that (right column) with the trigger only for the jet ($p_{T,\rm jet} > 80$ GeV/$c$) as shown in Fig. \ref{fig:map1}. 
    }
    \scalebox{0.89}[0.89]{
\begin{tabular}{ lcc}
\hline \hline
$p_{T}$ range for jet trigger (GeV/$c$)& 
 With $\gamma$ trigger &  Without $\gamma$ trigger \\
\hline
$110 < p_{T,\rm jet} < 120$   & 0.06&0.012\\
$100 < p_{T,\rm jet} < 110$& 0.26&0.047\\
$90 < p_{T,\rm jet} < 100$& 0.47&0.085\\
$80 < p_{T,\rm jet} < 90$ &0.21&0.037\\
\hline \hline
\end{tabular}
    \label{tab}
    }
  \end{center}
\end{table}

\begin{figure}
\begin{center}
 \includegraphics[width=8.0cm,bb=0 0 1024 768]{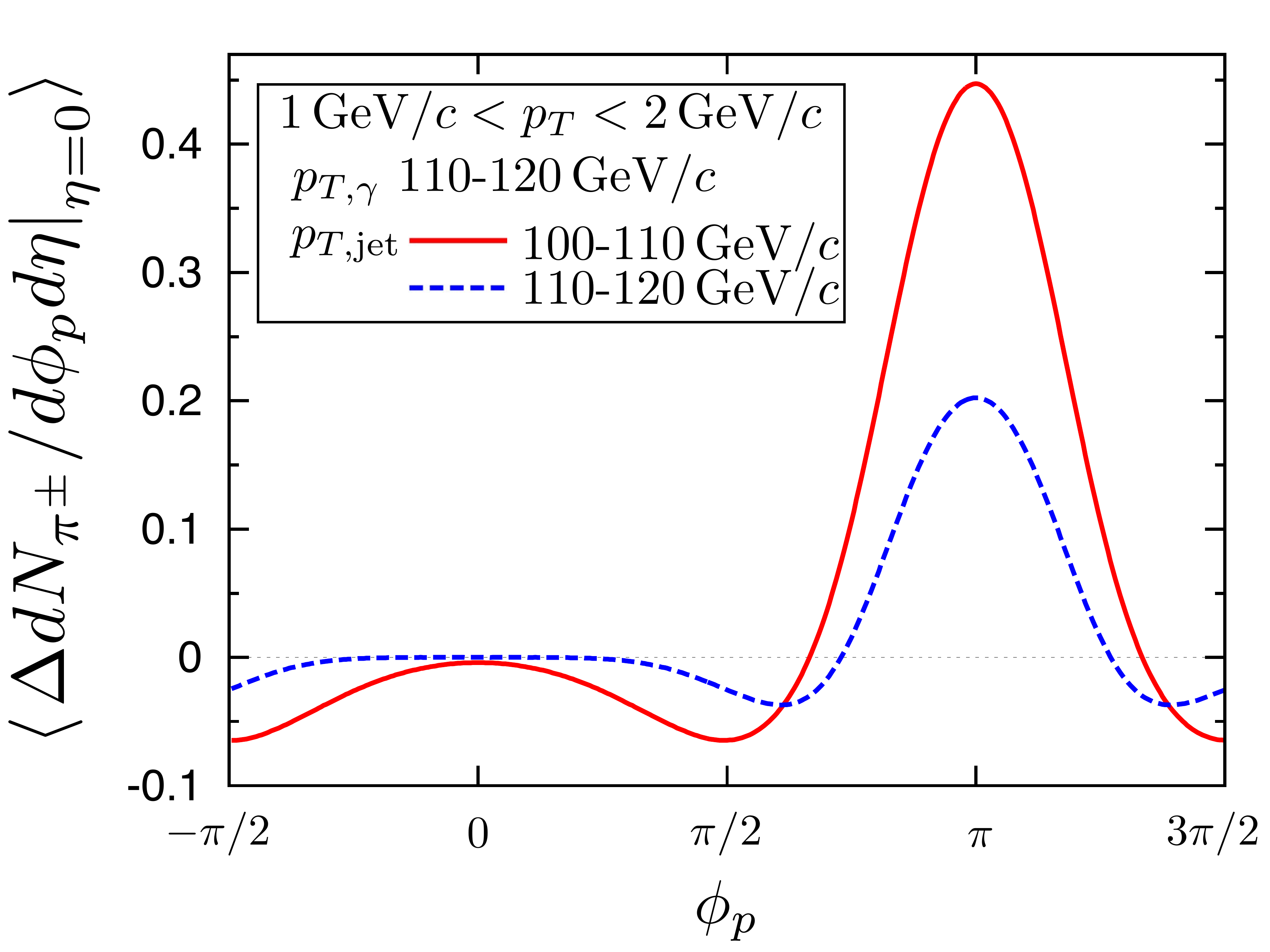} 
    \caption{(Color online)
      Azimuthal angle distributions 
      of charged pions with $1<p_{T}<2\,{\rm GeV}/c$ 
      subtracted by the background. 
      The trigger range of 
      the photon transverse momentum is 
      $110$-$120\,{\rm GeV}/c$. 
      The trigger ranges for 
      jet transverse momentum at final state 
      are 
      $100$-$110\,{\rm GeV}/c$ (solid line) 
      and 
      $110$-$120\,{\rm GeV}/c$ (dashed line) 
    }
    \label{fig:sp34}
    \end{center}
  \end{figure}

\begin{figure}
\begin{center}
 \includegraphics[width=8.0cm,bb=0 0 1024 768]{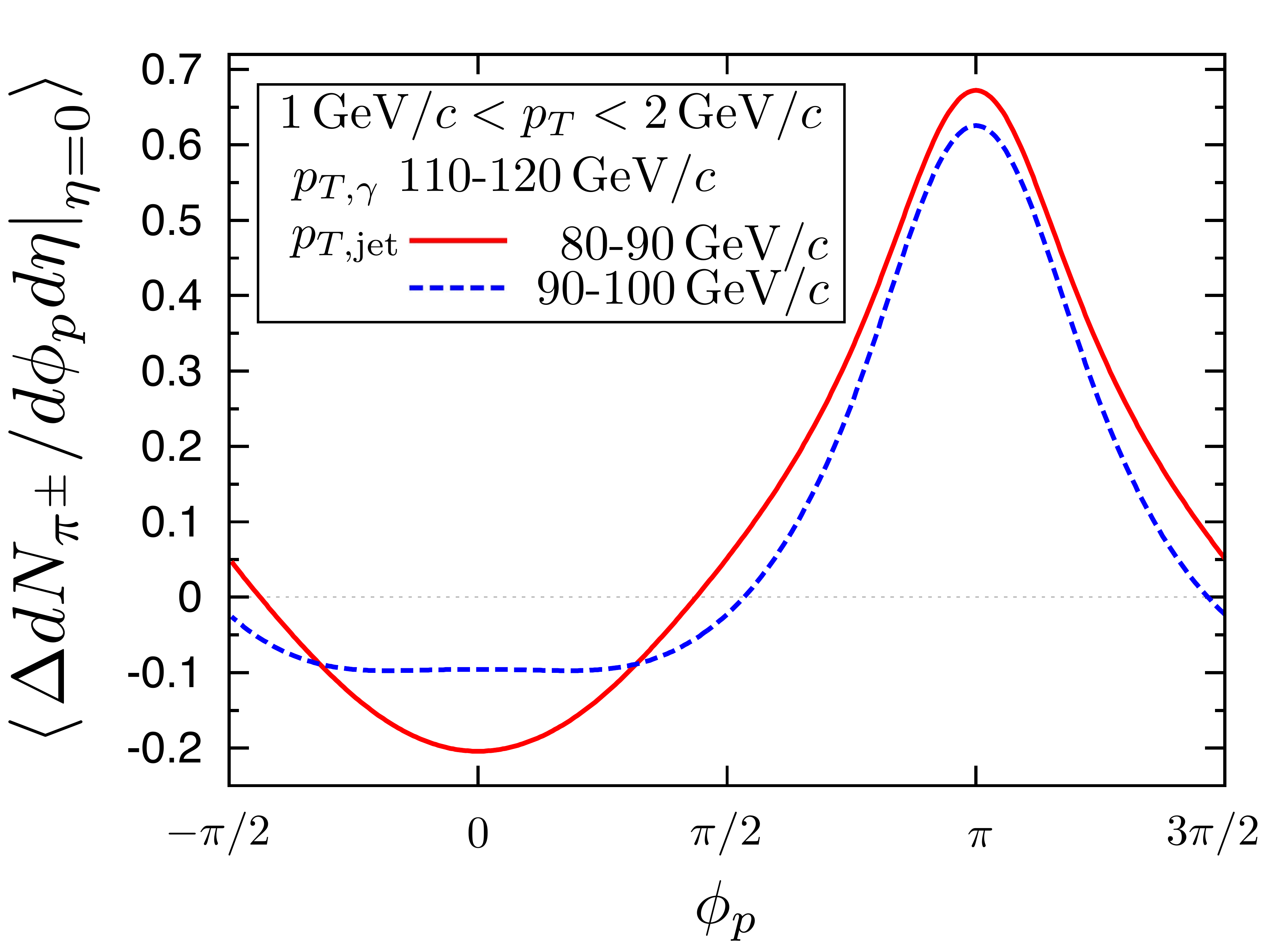} 
    \caption{(Color online)
      Azimuthal angle distributions 
      of charged pions with $1<p_{T}<2\,{\rm GeV}/c$ 
      subtracted by the background. 
      The trigger range of 
      the photon transverse momentum is 
      $110$-$120\,{\rm GeV}/c$. 
      The trigger ranges for 
      jet transverse momentum at final state 
      are 
      $80$-$90\,{\rm GeV}/c$ (solid line) 
      and 
      $90$-$100\,{\rm GeV}/c$ (dashed line).
    }
    \label{fig:sp12}
    \end{center}
  \end{figure}

Shown in Fig. \ref{fig:sp34} 
are the event-averaged 
azimuthal angle distributions 
at midrapidity 
when 
the $p_{T,\,\rm jet}$ trigger 
is 
$100$-$110$ GeV$/c$ 
and 
$110$-$120$ GeV$/c$. 
The trigger for $p_{T,\gamma}$ 
is also set to 
$110$-$120$ GeV$/c$. 
Two dips 
can be seen 
at both ends of 
the peak 
in the jet direction. 
This structure 
reflects 
the development 
of the Mach cone 
in the expanding medium 
and 
the crescent shape 
of the jet production point distribution. 
When a jet parton is produced  
at the edge of the crescent, 
the wave front
of the Mach cone 
on the center side 
pushes back 
the radial flow significantly. 
The resulting spectrum 
for such an event 
has a dip 
in the direction 
in which 
the jet path lies. 
The dips around $\phi=\pi/2$ 
are due to 
the contribution of 
the events with a jet created in $y\geq0$ 
and 
the dips around $\phi=3\pi/2$ 
are due to the contribution of 
the events with a jet created in $y\leq0$. 
When a jet is created 
at the center of the crescent, 
the Mach cone 
is pushed 
mainly from the inside 
by the radial flow. 
In such an event, 
the spectrum 
is almost flat 
except for 
the peak 
in the jet direction 
as shown by the solid line 
in Fig.~\ref{fig:x-3}. 
Therefore, 
the dips 
are not 
smeared out 
by the contribution of 
such events 
and 
can be clearly seen 
even 
after taking the event average 
owing to the trigger bias. 
Figure \ref{fig:sp12} 
shows the event-averaged 
azimuthal angle distributions 
at midrapidity 
when 
the $p_{T,\,\rm jet}$ triggers 
are 
$90$-$100$ GeV$/c$ 
and 
$80$-$90$ GeV$/c$. 
In these $p_{T,\,\rm jet}$ ranges 
of the jet trigger, 
the dips 
at both ends of the peaks 
cannot be 
seen. 
As seen in Fig.~\ref{fig:map} (c) and (d), 
the distribution 
of the jet production point 
is concentrated 
so that 
the jets take 
long paths 
through 
the central region of 
the QGP medium. 
Thus, 
the contribution 
of the events 
whose resulting spectra 
are similar to 
the solid line 
in Figs. \ref{fig:x3} and \ref{fig:x0} 
becomes dominant 
and 
the azimuthal angle distributions 
have simple structures. 

In this analysis, we set 
the initial energy of the jet and the photon to be the same. 
The events were divided 
according to the jet production point 
by introducing trigger thresholds 
both for the jet and for the photon.  
If one generates the photon-jet pair 
with a realistic $x_T$ distribution, 
one can also use the $x_T$ 
to control the jet production point 
because events 
with large $x_T$ close to unity 
are dominated by 
events with small energy loss. 
Indeed, the distribution 
for small-energy-loss events 
[Figs.~\ref{fig:map} (a) and (b)] 
are similar to the one 
for large-$x_T$ events 
of the event-by-event calculations 
in Ref. \cite{Qin:2012gp}. 

\section{Summary}\label{sec:sum}
In this paper,
we studied 
the hydrodynamic response of QGP 
to the energy-momentum deposition from jets. 
We formulated the model 
including the ideal hydrodynamic equations 
to describe the space-time evolution of the medium. 
To consider 
the contribution 
of the incoming energy and momentum from jets,  
the source terms 
were introduced 
to the hydrodynamic equations. 
In this process, 
we assumed that 
the deposited energy and momentum 
are instantaneously thermalized 
in the QGP fluid.

We performed simulations 
of$\gamma$-jet events 
in central Pb-Pb collisions at LHC. 
A massless jet particle 
travels through 
the expanding QGP fluid 
while depositing 
its energy and momentum 
into the fluid. 
Since the jet moves 
faster than 
the sound velocity of the medium,
the Mach cone is induced 
as a hydrodynamic response 
to the jet propagation. 
The shape of the Mach cone 
is distorted by the background expansion 
and 
this affects significantly 
the resulting particle distribution. 
Especially 
when the jet path 
is off central in the medium, 
the distortion is 
manifestly asymmetric. 

Then 
how the hydrodynamic response 
to the jet quenching 
can be reflected 
in the resulting particle spectra 
in heavy-ion collisions 
was studied. 
We calculated 
the event-averaged azimuthal angle distribution 
of charged pions 
emitted from the medium 
after the hydrodynamic evolution. 
When 
the trigger threshold 
is set only for 
the transverse momentum 
of the jet in the final state, 
only a peak 
can be seen 
in the direction of the jet.
However, 
when 
the jet production point 
is restricted 
in the upper half plane 
or 
in the lower half plane 
in the transverse plane, 
there is a dip 
on the side 
on which 
the jet path lies 
in the medium. 
The dip 
can also be seen 
in the azimuthal angle distribution 
for a single event 
when 
the jet path is away from 
the center of the medium. 
The wave front of the Mach cone 
holds back the radial flow 
and reduces 
the particle emission 
in the corresponding direction. 
The dip appears 
as a consequence of 
the interplay between the Mach cone 
and the radial expansion. 

We also studied 
the case where 
trigger thresholds 
are set 
both for the jet and for the photon. 
The dip 
appears 
according to 
the jet path in the medium. 
The path 
can be restricted 
by setting the trigger threshold 
to extract the events 
with 
a specific amount of the energy loss. 
When the trigger threshold 
is set to extract 
small energy-loss events, 
two dips 
can be seen 
at both ends of 
the peak 
in the jet direction. 
The origin of the dips 
is the same as 
the previous one 
in a single event 
where the jet path is away from 
the center. 
These are the consequences of 
the interplay between 
the Mach cone 
and 
the radial expansion. 
When 
large energy-loss events 
are extracted, 
the dips cannot be seen 
because 
the contribution of the events 
that make the dips 
becomes less dominant. 

The dips 
in the azimuthal angle distribution 
of soft particles in $\gamma$-jet events 
can be direct signals of hydrodynamic response to jet quenching 
and also include 
the information of the jet path in the medium: 
which side in the transverse plane the jet path lies on 
and 
how long the jets travel through the QGP. 
The geometry 
and the radial flow 
of the medium 
play important roles 
in the formation of the dips 
and 
are determined by 
the initial condition 
of the medium profile. 
In this work, we employed 
the smooth initial profile 
calculated from 
the optical Glauber model 
and 
only considered the case 
of impact parameter $b = 0$. 
In this case, 
the anisotropic flow is driven 
only by jet propagation. 
However, 
the fluctuation 
of the initial condition 
can always cause the anisotropy 
and 
affect the geometry 
and radial flow. 
We would like to defer 
the event-by-event studies 
including 
the initial fluctuation effect 
as a future work.

\begin{acknowledgements}
The authors are grateful to 
G.-Y. Qin and X.-N. Wang for useful comments. 
The authors would also like to thank Y. Hirono 
for helpful discussions regarding numerical implementations.
Y. T. was supported by a JSPS Research Fellowship for Young Scientists 
and by an Advanced Leading Graduate Course for Photon Science grant. 
The work was supported by JSPS KAKENHI Grants No. 
13J02554 (Y.~T.) and No. 5400269 (T.~H.). 
\end{acknowledgements}

\bibliography{ref.bib}

\end{document}